\documentclass[a4paper,12pt]{article}
\usepackage{amsmath,amssymb}

\begin{document}

\allowdisplaybreaks{

\begin{titlepage}

\setcounter{page}{0}
\mbox{}
\begin{flushright}
OU-HET 403\\
hep-th/0112083
\end{flushright}
\begin{center}
\vspace{20mm}
{\Large\bf 
Gravity/Non-Commutative Yang-Mills Correspondence
and Doubletons
}

\vspace{30mm}

{\large
Dan Tomino\footnote{e-mail address: dan@het.phys.sci.osaka-u.ac.jp}} 

\vspace{15mm}

{\em Department of Physics, Osaka University,
Toyonaka, Osaka 560-0043, Japan}

\end{center}

\vspace{10mm}

\centerline{{\bf{Abstract}}}

\vspace{5mm}
We discuss the gravity dual description for 
a non-commutative Yang-Mills theory,
which reduces to that 
on $AdS_{5}\times S_{5}$ in the commutative limit.
It is found that
doubletons do not decouple 
in this dual gravity description
unless one takes the commutative limit.
The decoupling of the doubletons
in $AdS_{5}$ space implies that
the dual gauge theory has $SU(N)$ gauge symmetry.
Our result implies that
this gravity description 
is dual to non-commutative $U(N)$ gauge theory.
It is compatible with the claim that 
$U(1)$ and $SU(N)$ gauge symmetries can not separate
in non-commutative $U(N)$ gauge theory.
\end{titlepage}

\newpage
\section{Introduction}
In the last few years, it has become well known  
that the (super) Yang-Mills theory on non-commutative space
$[\hat{x^{\mu}},\hat{x^{\nu}}]=i\theta^{\mu\nu}$
is realized as a low-energy effective theory on 
the Dp-branes in the constant background of 
NS-NS $B$-field.
By quantization of open strings 
in the constant B-field background, 
the coordinates of
Dp-brane's worldvolume 
become non-commutative operators. 
Then the effective theory,
which consists of
massless excitation modes of open strings 
whose edges are on the Dp-branes, 
 become the super Yang-Mills theory 
on the non-commutative space \cite{SW}.

By such construction, 
one can expect D-brane would be a useful tool to study 
non-perturbative properties of 
non-commutative Yang-Mills, 
as commutative cases. 
From this point of view, 
gravity dual descriptions for 
the non-commutative Yang-Mills 
are suggested in \cite{MR,HI}.
They can be thought as ones of natural generalizations of
the AdS/CFT correspondence \cite{Mal}.
This gravity descriptions are given as 
continuous deformation from $AdS$ background by 
constants which parameterize non-commutativity
in dual theories.  
Therefore one may use the procedure
in AdS/CFT correspondence
\cite{W,GKP},
to obtain information of the non-commutative Yang-Mills 
in the strong coupling region
from weakly coupled supergravity.

The effective theory of Dp-branes 
was thought to be $U(N)$ Yang-Mills. 
In AdS/CFT correspondence, however,  
the gravity on $AdS$ space is dual to a $SU(N)$ Yang-Mills.
This was pointed out by Witten at early stage
\cite{W,AGMOO}.
The $U(1)$ part of $U(N)\simeq U(1)\times SU(N)$ 
corresponds to 
the center of mass motion of $N$
Dp-brane system.
Supergravity fields, 
which couple to gauge theory operators associated with the $U(1)$ part
on the $AdS$ boundary,
are known as doubletons.
\footnote{Doubletons form 
the smallest unitary irreducible representation
of supergroup $U(2,2|4)$.
Any Kaluza-Klein modes of type IIB supergravity 
compactified on $AdS_{5}\times S_{5}$ 
falls into unitary irreducible representation of $U(2,2|4)$
\cite{GM,GMZ}.}
Doubletons have a strange property.
They are pure gauge mode in the interior of the boundary, 
and decouple from physical modes 
of gravity and other fields 
\cite{KRN,FF}.
Interactions are purely localized on the boundary.       
To obtain the gravity dual description of 
the $U(N)$ Yang-Mills,
one needs to include doubleton degree of freedom 
on the boundary.
\footnote{
Recently, \cite{MMS} discussed this point in Appendix B.   
}

In the gravity dual description for 
non-commutative Yang-Mills, however,
this 5-dimensional gravity theory 
in the interior of the boundary
should be dual to the $U(N)$   
non-commutative Yang-Mills itself. 
Unlike commutative $U(N)$ Yang-Mills,
$U(1)$ and $SU(N)$ gauge symmetries cannot be separated
in the $U(N)$ non-commutative Yang-Mills theory
in consequence of the non-commutative $*$-product
\cite{Ar}.
For this reason, we naturally expect 
the 5-dimensional gravity
would contain physical modes of doubletons
in the interior of the boundary
and the dual theory would be a $U(N)$ 
non-commutative Yang-Mills theory.
If it is true, 
we have the following question:
how is the decoupling of doubletons prevented?
In this paper, we answer this question.\\

A original motivation for our subject comes from
arguments of non-commutative gauge dynamics
using brane configuration with NS-NS $B$-field
\cite{SJ,AMT,OT}.
If there is no $B$-field, 
the $U(1)$ part also decouple from
other $SU(N)$ part. 
If this is also true, it would be a trouble for 
the 4d ${\cal N}=2$ non-commutative theory.
In the ordinary (commutative) case, 
it is known that
the $U(1)$ gauge symmetry is ``frozen out'' and
the theory constructed by branes should be
4d $\,{\cal N}=2\,$ $SU(N)$ gauge theory \cite{Wit}. 
On the other hand, 
$U(1)$ and  $SU(N)$ gauge symmetry cannot be separated 
in non-commutative $U(N)$ theory.
Therefore, 
when we construct non-commutative gauge theories
by brane configuration with $B$-field,  
we should ask
whether the gauge symmetry is
$U(N)$ or $SU(N)$.

In gravity description, 
the separation of the $U(1)$ gauge symmetry 
corresponds to the decoupling of doubletons.
It is an advance on the brane configuration argument
to analyze our subject.
So we try to our subject along the viewpoint of
gravity description and doubletons.
\footnote{
In this paper, we examine
the gravity description dual to 
4d ${\cal N}=4$ theory.
4d ${\cal N}=2$ case is left for 
a future study.
}   
\\

The mass spectrum of type IIB supergravity 
compactified on $AdS_{5}\times S_{5}$ was
studied by Kim, Romans and van Nieuwenhuizen \cite{KRN}.
This theory gives (at low energy) 
the gravity dual description of the commutative
4d ${\cal N}=4$ Yang-Mills. 
It was argued that
doubletons can be gauged away and decouple
from the gravity and other fields,
at least in the linearized level.
In this paper, we compactify  type IIB supergravity on
${\cal M}_{5}\times S_{5}$ following  \cite{KRN},
where ${\cal M}_{5}$ is continuous deformation from
$AdS_{5}$.
This theory is (at low energy) 
the gravity dual description of  
non-commutative Yang-Mills.
We obtain 5-dimensional linearized equations
from the compactification. 

To find the modes of non-decoupling doubletons,
we compare our equations to those in 
\cite{KRN}.
Then we find two independent equations of scalar fields.
These two equations have 
two independent solutions 
unless the parameter of non-commutativity $a$ vanishes. 
But when this parameter vanishes, 
these two equations degenerate, 
and we obtain only one equation.
Then this equation has 
infinite number of solutions.
It is related to 
the appearance of local symmetry
in $a=0$, 
and modes in one of the two field equations 
can be gauged away 
as gauge modes of the local symmetry. 
These modes are exactly
the scalars in doubleton multiplet. 
This is just what we expect.     \\

This paper is organized as follows : 
In section 2, we review the gravity dual description
of 
non-commutative Yang-Mills theory, which is our starting point.
In section 3, we consider linearized field equations in 
the gravity dual background, and
expand these equations 
by 5-dimensional spherical harmonics.
In section 4, which is the main part of this paper, 
we discuss the field equations associated with 
doubleton scalars.
A short comment for doubleton 2-forms is also given.
Finally, in section 5, implication to the dual 
non-commutative Yang-Mills
is discussed. \\

Throughout the paper, 
we discuss the bosonic part only.
It is sufficient to investigate this part
for our purpose.

\section{Gravity dual description of non-commutative Yang-Mills
  theory}
Our starting point is
a gravity dual description of 
4d non-commutative Yang-Mills theory \cite{MR,HI}.
\footnote{
In this paper, we use the gravity solution 
given in \cite{MR}.
}
This is the string theory on the spacetime 
${\cal M}_{5}\times S_{5}$
with flux of external fields :
 
\begin{eqnarray}
ds_{str}^2&=&\alpha^{\prime} R^2
\Bigg[u^2(-dx_{0}^2+dx_{1}^2)
+\frac{u^2}{1+a^4u^4}(dx_{2}^2+dx_{3}^2)
+\frac{du^2}{u^2}+d\Omega_{S_{5}}^2\Bigg] \;,
\nonumber \\
B_{23}&=&B_{\infty}\;\frac{a^4u^4}{1+a^4u^4}, \qquad
B_{\infty}=\alpha^{\prime}\frac{R^2}{a^2} \;,
\nonumber \\
e^{2\Phi}&=&\frac{\hat{g}^2}{1+a^4u^4} \;,
\nonumber \\
A_{01}&=&\left(\frac{1}{\hat{g}}\frac{\alpha^{\prime}R^2}{a^2}\right)\,a^4u^4 
\;,
\nonumber \\
\tilde{F}_{0123u}&=&\left(\frac{1}{\hat{g}}\frac{(\alpha^{\prime}R^2)^2}{a^3}\right)\,\frac{4a^3u^3}{1+a^4u^4}.
\label{5/31/1}
\end{eqnarray}

This solution is written in string metric.
$\Phi$ is a dilaton, 
$B_{23}$ and $A_{01}$ are NS-NS and R-R 2-form fields,
and $\tilde{F}_{0123u}$ is a 
5-form field strength which satisfies
the self-duality condition.
These backgrounds are obtained by taking decoupling 
and non-commutative scaling limit in
the supergravity solution of D3-brane with 
the $B_{23}$ field.
$R$ is written by the scaled string coupling constant 
$\hat{g}$ and the number of D3-branes $N$ ; 
$R^4=4\pi\hat{g}N$.

The parameter $a$ represents non-commutativity.
Non-commutativity appears in $x^{2},x^{3}$ directions 
$[x^{2},x^{3}]=\left(\frac{1}{B_{\infty}}\right)
\sim a^2$  
in the dual theory,
because $B_{23}$ has a non-zero value  
in solution (\ref{5/31/1}).

For $au\rightarrow 0$, 
the gravity solution (\ref{5/31/1}) becomes 
$AdS_{5}\times S_{5}$ solution. 
This means that 
the non-commutative theory reduces to
commutative one in the IR scale.

\section{Spherical harmonics expansion of bosonic field equations}
Now we expand bosonic field equations of type IIB supergravity
in small fluctuations around 
the solution (\ref{5/31/1})
in the previous section.
Making this expansion to first order in the fluctuations,
one finds linearized field equations. 
From now on, we use
the notation and convention 
which are given in Appendix A.

Bosonic field equations in the string metric are  
\begin{eqnarray}
\nabla^{M}\nabla_{M}\;\Phi
&=&\nabla^{M}\Phi\nabla_{M}\Phi
-\frac{1}{4}R
+\frac{1}{48}H^{MNP}H_{MNP} \;,
\label{3/18/1} \\
\nonumber \\
\nabla^{M}\nabla_{M}\;C
&=&-\frac{1}{3!}(F^{MNP}-CH^{MNP})H_{MNP} \;,
\label{3/20/1} \\
\nabla^{P}\left[F_{PMN}-CH_{PMN}\right]
&=&\frac{1}{3!}\tilde{F}_{MN}^{\quad \;PQR}H_{PQR} \;,
\label{3/18/3} \\
\nabla^{P}(e^{-2\Phi}H_{PMN})
&=&-\nabla^{P}\big[C(F_{PMN}-CH_{PMN})\big]
-\frac{1}{3!}\tilde{F}_{MN}^{\quad \;PQR}F_{PQR} \;,
\label{3/18/4} \\
R_{MN}&=&
-2\nabla_{M}\nabla_{N}\Phi
+\frac{1}{2}e^{2\Phi}\nabla_{M}C\nabla_{N}C
\nonumber \\
&&
+\frac{1}{4}H_{MPQ}H_{N}^{\;\;\;PQ}
+\frac{1}{4}e^{2\Phi}(F-CH)_{MPQ}(F-CH)_{N}^{\;\;\;PQ}
\nonumber \\
&&+\frac{1}{4\cdot4!}e^{2\Phi}\tilde{F}_{MPQRS}\tilde{F}_{N}^{\;\;PQRS} 
-g_{MN}\Bigg[
\frac{1}{4}\nabla^{M}\nabla_{M}\Phi
-\frac{1}{2}\nabla^{M}\Phi\nabla_{M}\Phi
\nonumber \\
&&+\frac{1}{48}H^{MNP}H_{MNP}
+\frac{1}{48}e^{2\Phi}(F-CH)^{MNP}(F-CH)_{MNP}
\Bigg],
\label{3/18/5} \\
\nonumber \\
\tilde{F}_{MNPQR}&=&\frac{1}{5!}\;\epsilon_{MNQPRSTUVW}\;
\tilde{F}^{STIVW}\;,
\label{5/31/2}
\end{eqnarray}
where, $C$ is R-R scalar.
$H^{(3)}=dB^{(2)}$ and $F^{(3)}=dA^{(2)}$ are
NS-NS and R-R 3-form field strength, respectively.
$\tilde{F}^{(5)}$ is defined by
$\tilde{F}^{(5)}=dD^{(4)}-\frac{1}{2}A^{(2)}\wedge H^{(3)}
+\frac{1}{2}B^{(2)}\wedge F^{(3)}$, and
$D^{(4)}$ is R-R 4-form. \\

We consider small fluctuations around 
the gravity solution (\ref{5/31/1}):
\begin{eqnarray}
\Phi&\rightarrow& \Phi+\phi 
, \qquad\qquad C\rightarrow 0 + c \; ,
\nonumber \\
B^{(2)}&\rightarrow&B^{(2)} + {\cal B}^{(2)}, \qquad 
A^{(2)}\rightarrow A^{(2)} + {\cal A}^{(2)} \;, 
\nonumber \\
D^{(4)}&\rightarrow& D^{(4)}+{\cal D}^{(4)} \;,
\,\qquad 
g_{\mu\nu}\rightarrow g_{\mu\nu}+h_{\mu\nu}
+\frac{1}{2}g_{\mu\nu}\phi \quad.
\label{12/1/1}
\end{eqnarray}

Mixing with metric fluctuation and dilaton 
comes from 
$g_{\mu\nu}^{(s)}=e^{\Phi/2}g_{\mu\nu}^{(E)}$.
This is the relation between
the string metric $g_{\mu\nu}^{(s)}$ and
the Einstein metric $g_{\mu\nu}^{(E)}$. 
\footnote{In \cite{KRN}, field equations 
written in the Einstein 
metric was used.
Due to linearized Weyl shift, we can compare
our results with those in \cite{KRN} at $a=0$.}
Substituting (\ref{12/1/1}) into field equations
(\ref{3/18/1})-(\ref{5/31/2}), 
one obtains linearized 
field equations.
All results are given in Appendix B. \\

Next, following the procedure in \cite{KRN}, 
let us expand linearized field equations by 
5-dimensional spherical harmonics.
Then we obtain 5-dimensional equations which
depend on coordinates 
$x^{0},x^{1},x^{2},x^{3},u$ only.
When $a\rightarrow 0$, these results reduce to
those of $AdS_{5}\times S_{5}$ compactification 
of type IIB supergravity
discussed by Kim, Romans and van Nieuwenhuizen.

Spherical harmonics expansions are as follows :
\begin{eqnarray}
&&\phi=\sum \phi^{I_{1}}(x)Y^{I_{1}}(y) \;, \quad 
c=\sum  c^{I_{1}}(x)Y^{I_{1}}(y) \;, 
\label{3/25/2} \nonumber \\
\nonumber \\
&&h_{\mu\nu}=h^{\prime}_{\mu\nu}-\frac{1}{3}g_{\mu\nu}\;h^{m}_{\;\;m}, \quad 
h_{mn}=h_{(mn)}+\frac{1}{5}\;g_{mn}\;h^{l}_{\;\;l} \;,
\nonumber \\
\nonumber \\
&&h^{\prime}_{\mu\nu}=\sum h^{I_{1}}_{\mu\nu}(x)Y^{I_{1}}(y),\quad 
h^{m}_{\;\;m}=\sum \pi^{I_{1}}(x)\;Y^{I_{1}}(y) \;,
\nonumber \\
&&h_{\mu m}=\sum h^{I_{5}}_{\mu}(x)Y_{m}^{I_{5}}(y)
, \quad 
h_{(mn)}=\sum h^{I_{14}}(x)Y_{(mn)}^{I_{14}}(y) \;,
\label{3/25/3} \nonumber \\
\nonumber \\
&&{\cal B}_{\mu\nu}=\sum b^{I_{1}}_{\mu\nu}(x)Y^{I_{1}}(y) 
,\quad 
{\cal A}_{\mu\nu}=\sum a^{I_{1}}_{\mu\nu}(x)Y^{I_{1}}(y)
\;, \nonumber \\
&&{\cal B}_{\mu m}=\sum b^{I_{5}}_{\mu}(x)Y_{m}^{I_{5}}(y) ,\quad
{\cal A}_{\mu m}=\sum a^{I_{5}}_{\mu}(x)Y_{m}^{I_{5}}(y) 
\;,\nonumber \\
&&{\cal B}_{mn}=\sum b^{I_{10}}(x)\;Y_{[mn]}^{I_{10}}(y) ,\quad
{\cal A}_{mn}=\sum a^{I_{10}}(x)\;Y_{[mn]}^{I_{10}}(y) 
\;,\nonumber \\
\label{3/25/4}
\nonumber \\
&&{\cal D}_{\mu\nu\rho\sigma}=\sum
d^{I_{1}}_{\mu\nu\rho\sigma}(x)Y^{I_{1}}(y) \;,\nonumber \\
&&{\cal D}_{\mu\nu\rho\,m}=\sum
d^{I_{5}}_{\mu\nu\rho}(x)\;Y_{m}^{I_{5}}(y) ,\quad
{\cal D}_{\mu\nu\,mn}=\sum
d^{I_{10}}_{\mu\nu}(x)\;Y_{[mn]}^{I_{10}}(y) \;,
\nonumber \\
&&{\cal D}_{\mu\,mnl}=\sum
d^{I_{5}}_{\mu}(x)\;\epsilon_{mnl}^{\quad
  \;\;ab}\nabla_{a}Y_{b}^{I_{5}}(y) ,\quad
{\cal D}_{mnlk}=\sum
d^{I_{1}}(x)\;\epsilon_{mnlk}^{\quad\;\;\;a}\nabla_{a}Y^{I_{1}}(y) \;,
\label{12/1/2} 
\end{eqnarray}
where $Y$'s are 5-dimensional spherical harmonics 
(see Appendix A).

These expansions are chosen to satisfy gauge conditions
\begin{eqnarray*}
\nabla^{m}h_{\mu m }=\nabla^{m}h_{(mn)}=0 \;,
\label{9/23/1} \\
\nabla^{m}{\cal B}_{\mu m}=\nabla^{m}{\cal B}_{mn}=0 \;,
\label{9/23/2} \\
\nabla^{m}{\cal A}_{\mu m}=\nabla^{m}{\cal A}_{mn}=0 \;,
\label{9/23/3} \\
\nabla^{m}{\cal D}_{\mu\nu\rho\,m}=\nabla^{m}{\cal
  D}_{\mu\nu\,mn}=\nabla^{m}{\cal D}_{\mu\,mnl}=\nabla^{m}{\cal
  D}_{mnlk}=0 \;.
\label{9/23/4}
\end{eqnarray*}
These conditions respect the invariance under
(i) 5-dimensional diffeomorphism;  
$\delta h_{\mu\nu}=\nabla_{\mu}\xi_{\nu}(x)
+\nabla_{\nu}\xi_{\mu}(x)$,
(ii) Yang-Mills symmetries for which 
$\xi_{m}=\sum\lambda(x)^{I}Y^{I}_{m}$
with $Y^{I}_{m}$ equal to Killing vectors,
(iii) conformal diffeomorphism
for which 
$\xi_{m}=\sum\kappa^{I}(x)\nabla_{m}Y^{I}(y),\;\xi_{\mu}=
-\sum\nabla_{\mu}\kappa^{I}(x)Y^{I}(y)$.
Those $Y^{I}(y)$ satisfy
$\nabla_{(m}\nabla_{n)}Y^{I}=0$, 
which are called conformal scalars in \cite{Nie}. \\

Substituting expansions (\ref{12/1/2}) into
linearized equations in Appendix B, 
we obtain 5-dimensional field equations on ${\cal M}_{5}$.
All results are given in Appendix C.

\section{Doubleton non-decoupling}
The doubleton multiplet is 
$(0,0,\mbox{\boldmath $6$})\oplus
(\frac{1}{2},0,\mbox{\boldmath $4$})\oplus
(0,\frac{1}{2},\mbox{\boldmath $\bar{4}$})\oplus
(1,0,\mbox{\boldmath $1$})\oplus
(0,1,\mbox{\boldmath $1$})$ of
$(SU(2)_{A},SU(2)_{B},SO(6))$ ;
these groups are compact subgroups of $U(2,2|4)$.
In harmonic expansion (\ref{12/1/2}),
the modes associated with doubletons 
appear as coefficient of
scalar spherical harmonics $Y^{I_{1}}$ and 
their derivatives.
So, in this section, we concentrate on 
the equations in such a class.

They are the following six equations : 
\begin{eqnarray}
&\Bigg\{&\frac{1}{2}(2\nabla_{x}^2+\nabla_{y}^2)h^{I_{1}\;\rho}_{\rho}
-\nabla^{\mu}\nabla^{\nu}h^{I_{1}}_{\mu\nu}
+\left(-\frac{7}{120}(H^2+e^{2\Phi}F^2)
-\frac{1}{5!}e^{2\Phi}\tilde{F}_{\mu\nu\rho\sigma\tau}
\tilde{F}^{\mu\nu\rho\sigma\tau}
\right)h^{I_{1}\;\rho}_{\rho}
\nonumber \\
&&\hspace{-5mm}
-\frac{5}{6}(\nabla_{x}^2+\nabla_{y}^2)\phi^{I_{1}}
+\left(-\frac{5}{3}\left[\frac{1}{4}\nabla_{x}^2\Phi
-\frac{1}{2}(\nabla\Phi)^2\right]
+\frac{1}{16}(H^2+e^{2\Phi}F^2)
+\frac{1}{3\cdot4!}e^{2\Phi}\tilde{F}_{\mu\nu\rho\sigma\tau}
\tilde{F}^{\mu\nu\rho\sigma\tau}
\right)\pi^{I_{1}}
\nonumber \\
&&+\left(5\;(\frac{1}{4}\nabla^{\rho}\nabla^{\sigma}\Phi
-\frac{1}{2}\nabla^{\rho}\Phi\nabla^{\sigma}\Phi)
-\frac{3}{16}(H_{\mu\nu}^{\;\;\;\rho}H^{\mu\nu\sigma}
+e^{2\Phi}F_{\mu\nu}^{\;\;\;\rho}F^{\mu\nu\sigma})
+\frac{1}{4!}e^{2\Phi}
\tilde{F}_{\mu\nu\lambda\tau}^{\quad\;\;\rho}
\tilde{F}^{\mu\nu\lambda\tau\sigma}
\right)h^{I_{1}}_{(\rho\sigma)}
\nonumber \\
&&-\frac{13}{4}\nabla_{x}^2\phi^{I_{1}}
+5\nabla\Phi\nabla\phi^{I_{1}}
+\left(-\frac{5}{4}\nabla_{y}^2
+\frac{1}{24}e^{2\Phi}F^2
+\frac{1}{2\cdot4!}e^{2\Phi}
\tilde{F}_{\mu\nu\rho\sigma\tau}
\tilde{F}^{\mu\nu\rho\sigma\tau}
\right)\phi^{I_{1}}
\nonumber \\
&&\hspace{-10mm}
-\frac{5}{4}\nabla^{\mu}\Phi
\left(\nabla^{\nu}h^{I_{1}}_{(\mu\nu)}
-\frac{3}{10}\nabla_{\mu}h^{I_{1}\;\rho}_{\rho}
\right)
+\frac{7}{24}(H{\cal H}+e^{2\Phi}F{\cal F})
+\frac{1}{2\cdot4!}e^{2\Phi}
\tilde{F}^{\mu\nu\rho\sigma\tau}
{\cal \tilde{F}}^{I_{1}}_{\mu\nu\rho\sigma\tau}
\Bigg\}\;Y^{I_{1}}
 = 0\;,
\label{trE1.1} 
\end{eqnarray}

\begin{eqnarray}
&\Bigg\{&\frac{1}{2}\nabla_{x}^2\pi^{I_{1}}
+\left(-\frac{1}{30}\nabla_{y}^2
-\frac{1}{5!}e^{2\Phi}\tilde{F}_{mnpqr}\tilde{F}^{mnpqr}
-\frac{2}{3}\nabla_{x}\Phi
+\frac{4}{3}(\nabla\Phi)^2
-\frac{1}{8}(H^2+e^{2\Phi}F^2)
\right)\pi^{I_{1}}
\nonumber \\
&&+\left(\frac{1}{2}\nabla_{y}^2
+\frac{1}{4}\nabla_{x}^2\Phi
-\frac{1}{2}(\nabla\Phi)^2
+\frac{1}{16}(H^2+e^{2\Phi}F^2)
\right)h^{I_{1}\;\rho}_{\rho}
-\frac{3}{8}\nabla^{\mu}\Phi\nabla_{\mu}h^{I_{1}\;\rho}_{\rho}
\nonumber \\
&&+\frac{5}{4}\nabla^{\mu}\Phi\nabla^{\nu}h^{I_{1}}_{(\mu\nu)}
+5\left(\frac{1}{4}\nabla^{\rho}\nabla^{\sigma}\Phi
-\frac{1}{2}\nabla^{\rho}\Phi\nabla^{\sigma}\Phi
+\frac{1}{16}(H_{\mu\nu}^{\;\;\;\rho}H^{\mu\nu\sigma}
+e^{2\Phi}F_{\mu\nu}^{\;\;\;\rho}F^{\mu\nu\sigma})
\right)h^{I_{1}}_{(\rho\sigma)}
\nonumber \\
&&-\frac{5}{4}\nabla_{x}^2\phi^{_{1}}
+5\nabla\Phi\nabla\phi^{I_{1}}
+\left(-\frac{13}{4}\nabla_{y}^2
-\frac{5}{24}e^{2\Phi}F^2
+\frac{1}{2\cdot4!}e^{2\Phi}
\tilde{F}_{mnpqr}\tilde{F}^{mnpqr}
\right)\phi^{I_{1}}
\nonumber \\
&&-\frac{5}{24}(H{\cal H}+e^{2\Phi}F{\cal F})
+\frac{5}{2\cdot4!}e^{2\Phi}\tilde{F}^{mnpqr}
\nabla_{m}\epsilon_{npqr}^{\quad\;\;\;a}\nabla_{a}\;d^{I_{1}}
\Bigg\}\;Y^{I_{1}}\quad =\quad 0 \;.
\label{trE3.1}
\end{eqnarray}

\begin{eqnarray}
&\Bigg\{& \!\!\!
\frac{1}{2}\nabla_{\mu}h^{I_{1}\;\rho}_{\;\;\rho}
-\frac{1}{2}\nabla^{\rho}h^{I_{1}}_{\rho\mu}
-\frac{4}{15}\nabla_{\mu}\pi^{I_{1}}
+\frac{1}{4}(H_{\mu}^{\;\;\rho\sigma}b^{I_{1}}_{\rho\sigma}
+e^{2\Phi}F_{\mu}^{\;\;\rho\sigma}a^{I_{1}}_{\rho\sigma}
)
+\frac{1}{4\cdot4!}e^{2\Phi}\tilde{F}_{\mu}^{\;\;\nu\rho\sigma\tau}
\tilde{D}_{\nu\rho\sigma\tau}
\Bigg\}\;\nabla_{m}Y^{I_{1}}
\nonumber \\
&&+\frac{1}{4\cdot4!}\tilde{F}_{m}^{\;\;npqr}
\epsilon_{npqr}^{\quad\;\;a}\nabla_{\mu}d^{I_{1}}
\nabla_{a}Y^{I_{1}}
\quad =\quad 0 \;,
\label{10/10/1} 
\end{eqnarray}
\vspace{-5mm}
\begin{eqnarray}
\hspace{-70mm}
\Bigg\{\;\frac{1}{2}h^{I_{1}\;\rho}_{\;\;\rho}
-\frac{8}{15}\pi^{I_{1}}
\;\Bigg\}\;\nabla_{(m}\nabla_{n)}\;Y^{I_{1}}
\quad =\quad 0 
\label{10/10/4} \;, 
\end{eqnarray}

\begin{eqnarray}
&&\hspace{-13mm}
\Bigg\{
\tilde{{\cal F}}_{\mu\nu\rho\sigma\lambda}^{I_{1}}
-\epsilon_{\mu\nu\rho\sigma\lambda}
\;\nabla_{y}^2\;d^{I_{1}}
-\frac{1}{2\cdot5!}\;\epsilon_{\mu\nu\rho\sigma\lambda}
^{\qquad mnlkj}
\tilde{F}_{mnlkj}\left(
h_{\rho}^{I_{1}\;\;\rho}
-\frac{8}{3}\;\pi^{I_{1}}\right)\;
\Bigg \}\;Y^{I_{1}}
\quad = \quad 0 \;,
\label{11/29/1}  \\
\nonumber \\
&&\hspace{-13mm}
\Bigg\{
\tilde{D}_{\mu\nu\rho\sigma}^{I_{1}}
+\epsilon_{\mu\nu\rho\sigma\lambda}\;
\nabla^{\lambda}\;d^{I_{1}}
\Bigg\}\;\nabla_{m}Y^{I}(y)
\quad =\quad 0 \;,
\label{11/29/2}
\end{eqnarray}
Equations (\ref{trE1.1}) and (\ref{trE3.1}) are traces of
(\ref{E1}) and (\ref{E3.1}) in Appendix C respectively,  
which are derived from
linearized Einstein equations 
(\ref{Ein1}),(\ref{Ein3}) in Appendix B,
by harmonic expansions.
Equations (\ref{10/10/1}) and (\ref{10/10/4}) 
also come from (\ref{Ein1}) and (\ref{Ein3}). 
Equations (\ref{11/29/1}) and (\ref{11/29/2})
are derived from
the linearized self-dual equation (\ref{SD1})
in Appendix B,
by harmonic expansions.
Scalar harmonics $Y^{I_{1}}$ are eigenfunction 
of Laplacian $\nabla_{y}^2$
($=$ Hodge-de Rham operator $\Delta$ for $Y^{I_{1}}$)
with eigenvalues 
$-e^2k(k+4)$, where $k=0,1,2, ....$ 
and
$e^2=1/\alpha^{\prime}R^2$.
When $k=0$, $Y^{I_{1}}_{(k=0)}$ is constant and satisfies
$\nabla_{m}Y^{I_{1}}_{(k=0)}=0$. 
Then (\ref{10/10/1}),(\ref{10/10/4}) 
and (\ref{11/29/2}) 
become trivial (i.e. $0=0$).
When $k=1$, 
$Y^{I_{1}}_{(k=1)}$ are conformal scalars i.e.
$\nabla_{(m}\nabla_{n)}Y^{I}=0$. 
Then (\ref{10/10/4}) becomes trivial. \\

These equations contain 5-dimensional physical fields 
but unphysical fields are also contained. 
In the linearized level, these unphysical fields are
algebraically eliminated.

In fact,  the 5-form  
$\tilde{{\cal F}}^{I_{1}}_{\mu\nu\rho\sigma\tau}$ 
is algebraically eliminated by (\ref{11/29/1}),
for all $k\ge 0$. 
When $k\ge 1$, 
we obtain the field equation 
for scalar fields  $d^{I_{1}}$ :
\begin{eqnarray}
\Bigg\{
(\nabla_{x}^2+\nabla_{y}^2)d^{I_{1}}
+\frac{1}{5!}\;\epsilon^{mnlkj}\tilde{F}_{mnlkj}
\left(\frac{1}{2}h^{I_{1}\;\rho}_{\rho}
-\frac{4}{3}\pi^{I_{1}}\right)
+\frac{10}{5!}\;
\epsilon^{\mu\nu\rho\sigma\lambda}
(b^{I_{1}}_{\mu\nu}F_{\rho\sigma\lambda}
-a^{I_{1}}_{\mu\nu}H_{\rho\sigma\lambda})
\Bigg\}\;Y^{I_{1}} = 0 \;,
\label{9/14/2}
\end{eqnarray}
from a combination of (\ref{11/29/1}) and (\ref{11/29/2}). 

Before we discuss the decoupling of doubletons in detail, 
here let us give the outline of the decoupling.
Similarly to 
$\tilde{{\cal F}}^{I_{1}}_{\mu\nu\rho\sigma\tau}$,
we can algebraically eliminate 
the scalar fields $h^{I_{1}\;\rho}_{\rho}$ 
as unphysical modes
for $k\ge1$ ,
and obtain two independent modes 
associated with equations for scalar fields 
$\pi^{I_{1}}, d^{I_{1}}$.
This procedure breaks down when 
the non-commutative parameter $a$ vanishes.
At $a=0$, scalar field equations for $k=1$ 
have conformal diffeomorphism invariance.
Then the scalar modes associated with 
one of the two scalar field equations
become pure gauge modes
of the local gauge symmetry.
These pure gauge modes can be gauged away
and disappear from 5d equations.  
This is the decoupling of doubleton scalars.
\\

Now, we discuss the decoupling mechanism in more detail.

If $k\ge2$, we can eliminate 
scalar fields $h^{I_{1}\;\rho}_{\;\;\rho}$ 
from (\ref{trE3.1}),(\ref{9/14/2}) by using 
(\ref{10/10/4}). 
Then we obtain two independent field equations 
for scalar fields 
$\pi^{I_{1}}, d^{I_{1}}$ which have mixing with 
other tensors.\\

Next, we consider the case of $k=1$.
For $k=1$, (\ref{10/10/4}) is trivial.
To eliminate $h^{I_{1}\;\rho}_{\;\;\rho}$ from (\ref{trE3.1}),
we use (\ref{9/14/2}). 
Then one finds
\begin{eqnarray}
&&\frac{1}{2}\nabla_{x}^2\pi^{I_{1}}
+\left(\frac{13}{10}\nabla_{y}^2
-\frac{1}{5!}e^{2\Phi}\tilde{F}_{mnpqr}\tilde{F}^{mnpqr}
-\frac{1}{24}(H^2+e^{2\Phi}F^2)
\right)\pi^{I_{1}}
-\frac{1}{4}\nabla^{\mu}\Phi\nabla_{\mu}\pi^{I_{1}}
\nonumber \\
&&-\left(\nabla_{y}^2
+\frac{1}{2}\nabla_{x}^2\Phi
-(\nabla\Phi)^2
+\frac{1}{8}(H^2+e^{2\Phi}F^2)
\right)
\left(\frac{1}{5!}\tilde{F}^{mnpqr}\epsilon_{mnpqr}
\right)^{-1}
(\nabla_{x}^2+\nabla_{y}^2)\;d^{I_{1}}
\nonumber \\
&&+\frac{5}{2\cdot4!}e^{2\Phi}\tilde{F}^{mnpqr}
\nabla_{m}\epsilon_{npqr}^{\quad\;\;a}\nabla_{a}d^{I_{1}}
-\frac{15}{64\cdot4!}e^{2\Phi}(\nabla_{\mu}\Phi)
(\tilde{F}^{\mu\nu\rho\sigma\tau}
\epsilon_{\nu\rho\sigma\tau\lambda}
\nabla^{\lambda}d^{I_{1}}
-\tilde{F}^{mnpqr}\epsilon_{npqr a}\nabla_{\mu}d^{I_{1}})
\nonumber \\
&&+\frac{25}{32}\nabla^{\mu}\Phi\nabla^{\nu}h^{I_{1}}_{(\mu\nu)}
+5\left(\frac{1}{4}\nabla^{\rho}\nabla^{\sigma}\Phi
-\frac{1}{2}\nabla^{\rho}\Phi\nabla^{\sigma}\Phi
+\frac{1}{16}(H_{\mu\nu}^{\;\;\;\rho}H^{\mu\nu\sigma}
+e^{2\Phi}F_{\mu\nu}^{\;\;\;\rho}F^{\mu\nu\sigma})
\right)h^{I_{1}}_{(\rho\sigma)}
\nonumber \\
&&-\frac{5}{4}\nabla_{x}^2\phi^{_{1}}
+\frac{55}{8}\nabla\Phi\nabla\phi^{I_{1}}
+\left(-\frac{13}{4}\nabla_{y}^2
-\frac{5}{24}e^{2\Phi}F^2
+\frac{1}{2\cdot4!}e^{2\Phi}
\tilde{F}_{mnpqr}\tilde{F}^{mnpqr}
\right)\phi^{I_{1}}
\nonumber \\
&&-\frac{5}{24}(H{\cal H}+e^{2\Phi}F{\cal F})
+\frac{25}{64}\nabla^{\mu}\Phi
(H_{\mu}^{\;\;\rho\sigma}b^{I_{1}}_{\rho\sigma}
+e^{2\Phi}(F_{\mu}^{\;\;\rho\sigma}a^{I_{1}}_{\rho\sigma}))
\nonumber \\
&&\hspace{-20mm}
-\left(\nabla_{y}^2
+\frac{1}{2}\nabla_{x}^2\Phi
-(\nabla\Phi)^2
+\frac{1}{8}(H^2+e^{2\Phi}F^2)
\right)\!\!
\left(\frac{1}{5!}\tilde{F}^{mnpqr}\epsilon_{mnpqr}
\right)^{-1}\!\!\frac{10}{5!}\epsilon^{\mu\nu\rho\sigma\tau}
(F_{\rho\sigma\tau}b^{I_{1}}_{\mu\nu}
-H_{\rho\sigma\tau}a^{I_{1}}_{\mu\nu})
 =0 .
\label{E3.1b}
\end{eqnarray}
This is a field equation for 
$\pi^{I_{1}},d^{I_{1}}$ 
which have mixing with other tensors. 
Another field equation for 
$\pi^{I_{1}},d^{I_{1}}$
comes from (\ref{trE1.1}), 
which is independent of (\ref{E3.1b}).
We use (\ref{10/10/1}) and (\ref{9/14/2}) to
eliminate 
$\nabla_{\mu}\nabla_{\nu}h^{I_{1}\;\rho}_{\;\;\rho}$ 
and $h^{I_{1}\;\rho}_{\;\;\rho}$ from (\ref{trE1.1}).
Then one finds 
\begin{eqnarray}
&&\hspace{-20mm}
-\frac{3}{10}\nabla_{x}^2\pi^{I_{1}}
+\frac{1}{4}\nabla^{\mu}\Phi\nabla_{\mu}\pi^{I_{1}}
\!\!+\!\!\Bigg(\frac{1}{2}\nabla_{y}^2
-\frac{5}{3}\left[\frac{1}{4}\nabla_{x}^2\Phi
-\frac{1}{2}(\nabla\Phi)^2\right]
-\frac{67}{45\cdot16}(H^2+e^{2\Phi}F^2)
-\frac{e^{2\Phi}}{5!}\tilde{F}_{\mu\nu\rho\sigma\tau}
\tilde{F}^{\mu\nu\rho\sigma\tau}
\!\! \Bigg)\pi^{I_{1}}
\nonumber \\
&&
-\left(\nabla_{y}^2
-\frac{7}{60}(H^2+e^{2\Phi}F^2)
+\frac{1}{2\cdot5!}e^{2\Phi}\tilde{F}^{\mu\nu\rho\sigma\tau}
\tilde{F}_{\mu\nu\rho\sigma\tau}
\right)
\left(\frac{1}{5!}\tilde{F}^{mnpqr}\epsilon_{mnpqr}
\right)^{-1}
\;(\nabla_{x}^2+\nabla_{y}^2)\;d^{I_{1}}
\nonumber \\
&&-\frac{1}{2\cdot4!}e^{2\Phi}\nabla^{\mu}
(\tilde{F}_{m}^{\;\;npqr}\epsilon_{npqr}^{\quad\;a}
\nabla_{\mu}d^{I_{1}})
+\frac{1}{2\cdot4!}\nabla^{\mu}
(e^{2\Phi}\tilde{F}_{\mu}^{\;\;\nu\rho\sigma\tau}
\epsilon_{\nu\rho\sigma\tau\lambda}
\nabla^{\lambda}d^{I_{1}})
\nonumber \\
&&+\frac{1}{2\cdot4!}e^{2\Phi}
\tilde{F}^{\mu\nu\rho\sigma\tau}
\epsilon_{\mu\nu\rho\sigma\tau}
\nabla_{y}^2\;d^{I_{1}}
+\frac{15}{64\cdot4!}\nabla^{\mu}\Phi(e^{2\Phi}
\tilde{F}_{\mu}^{\;\;\nu\rho\sigma\tau}
\epsilon_{\nu\rho\sigma\lambda}\nabla^{\lambda}d^{I_{1}}
-e^{2\Phi}\tilde{F}^{mnpqr}
\epsilon_{npqr a}\nabla_{\mu}d^{I_{1}})
\nonumber \\
&&\hspace{-20mm}
-\frac{15}{32}\nabla^{\mu}\Phi\nabla^{\nu}h^{I_{1}}_{(\mu\nu)}
\!\!+\!\!
\left(5\;(\frac{1}{4}\nabla^{\rho}\nabla^{\sigma}\Phi
-\frac{1}{2}\nabla^{\rho}\Phi\nabla^{\sigma}\Phi)
-\frac{3}{16}(H_{\mu\nu}^{\;\;\;\rho}H^{\mu\nu\sigma}
+e^{2\Phi}F_{\mu\nu}^{\;\;\;\rho}F^{\mu\nu\sigma})
+\frac{1}{4!}e^{2\Phi}
\tilde{F}_{\mu\nu\lambda\tau}^{\quad\;\;\rho}
\tilde{F}^{\mu\nu\lambda\tau\sigma}
\!\!\right)h^{I_{1}}_{(\rho\sigma)}
\nonumber \\
&&+\frac{3}{4}\nabla_{x}^2\phi^{I_{1}}
+\frac{55}{8}\nabla\Phi\nabla\phi^{I_{1}}
+\left(-\frac{5}{4}\nabla_{y}^2
+\frac{1}{24}e^{2\Phi}F^2
+\frac{1}{2\cdot4!}e^{2\Phi}
\tilde{F}_{\mu\nu\rho\sigma\tau}
\tilde{F}^{\mu\nu\rho\sigma\tau}
\right)\phi^{I_{1}}
\nonumber \\
&&+\frac{1}{8}(H{\cal H}+e^{2\Phi}F{\cal F})
-\frac{79}{64}\nabla^{\mu}\Phi
(H_{\mu}^{\;\;\rho\sigma}b^{I_{1}}_{\rho\sigma}
+e^{2\Phi}F_{\mu}^{\;\;\rho\sigma}a^{I_{1}}_{\rho\sigma})
+\frac{1}{12}\tilde{F}^{\mu\nu\rho\sigma\tau}
(F_{\rho\sigma\tau}b^{I_{1}}_{\mu\nu}
-H_{\rho\sigma\tau}a^{I_{1}}_{\mu\nu})
\nonumber \\
&&\hspace{-25mm}
\!\!\!\!-\!\!\left(\nabla_{y}^2
-\frac{7}{60}(H^2+e^{2\Phi}F^2)
+\frac{e^{2\Phi}}{2\cdot5!}\tilde{F}^{\mu\nu\rho\sigma\tau}
\tilde{F}_{\mu\nu\rho\sigma\tau}
\right)\!\!\!
\left(\frac{1}{5!}\tilde{F}^{mnpqr}\epsilon_{mnpqr}
\right)^{-1}\!\!\!
\frac{10}{5!}\tilde{F}^{\mu\nu\rho\sigma\tau}
(F_{\rho\sigma\tau}b^{I_{1}}_{\mu\nu}
-H_{\rho\sigma\tau}a^{I_{1}}_{\mu\nu})
=0 \;. 
\label{E1.1b} 
\end{eqnarray}

Therefore, even if (\ref{10/10/4}) is trivial for $k=1$,
we can also eliminate $h^{I_{1}\;\rho}_{\;\;\rho}$
from field equations.
We can obtain two independent solutions for
$\pi^{I_{1}}, d^{I_{1}}$
by using (\ref{E3.1b}) and (\ref{E1.1b}). \\

However, if the non-commutative parameter $a$ becomes zero,
this procedure breaks down.
When $a=0$, equations (\ref{E3.1b}),(\ref{E1.1b}) 
degenerate to the same one : 
\begin{eqnarray}
\nabla_{x}^2(\pi^{I_{1}}+\frac{5}{2}ed^{I_{1}})
-45e^2(\pi^{I_{1}}+\frac{5}{2}ed^{I_{1}}) = 0\;. 
\label{10/13/1}
\end{eqnarray}
We then find that
(\ref{10/13/1}) has infinite number of solutions.
This is a consequence of 
conformal diffeomorphism invariance
\begin{eqnarray}
\pi^{I_{1}} \rightarrow \pi^{1_{1}}+10\lambda , \quad
ed^{I_{1}} \rightarrow ed^{1_{1}}-4\lambda \;,
\label{10/13/2}
\end{eqnarray}
for arbitrary function $\lambda(x)$.
The reason for these infinite number of solutions 
originates from the fact that
the local gauge symmetry (\ref{10/13/2}) is not fixed.

In this case, 
we should return to equations
(\ref{10/10/1}),(\ref{trE3.1}) and (\ref{9/14/2}). 
When $a=0$, these equations are
\begin{eqnarray}
\mbox{[eq.(\ref{10/10/1})]}&\Rightarrow&
\Bigg\{
\nabla^{\rho}h^{I_{1}}_{\rho\mu}
-\nabla^{\mu}(
h^{I_{1}\;\rho}_{\;\rho}
-\frac{8}{15}\pi^{I_{1}}
+4ed^{I_{1}})
\Bigg\}\;\nabla_{m}Y^{I_{1}} = 0 \;,
\label{11/25/1} \\
\mbox{[eq.(\ref{trE3.1})]}&\Rightarrow&
\Bigg\{
(\nabla_{x}^2+\nabla_{y}^2-32e^2)\;\pi^{I_{1}}
+20e\;\nabla_{y}^2d^{I_{1}}
+\nabla_{y}^2(h^{I_{1}\;\rho}_{\;\rho}
-\frac{16}{15}\pi^{I_{1}}) 
\Bigg\}\;Y^{I_{1}} = 0 \;,
\label{11/25/2} \\
\mbox{[eq.(\ref{9/14/2})]}&\Rightarrow&
\Bigg\{
(\nabla_{x}^2+\nabla_{y}^2)\;d^{I_{1}}
+4e(\frac{1}{2}h^{I_{1}\;\rho}_{\;\rho}
-\frac{4}{3}\pi^{I_{1}})
\Bigg\}\;Y^{I_{1}} = 0\;.
\label{11/25/3}
\end{eqnarray}
(\ref{trE1.1}) is linearly dependent on
(\ref{11/25/1}),(\ref{11/25/2}),(\ref{11/25/3})
and (\ref{10/10/4}) for $a=0$. 
And (\ref{10/10/4}) becomes trivial for $k=1$. 
So we do not have to consider
(\ref{trE1.1}),(\ref{10/10/4}) here.

Equations (\ref{11/25/1}),(\ref{11/25/2})
and (\ref{11/25/3}) are also invariant under 
conformal diffeomorphism : 
\begin{eqnarray}
h^{I_{1}\;\rho}_{\;\;\rho}\rightarrow 
h^{I_{1}\;\rho}_{\;\;\rho}+
\frac{2}{e}\nabla^2_
{x}\lambda+\frac{50}{3}\;\lambda, \quad 
\pi^{I_{1}} \rightarrow \pi^{I_{1}}+10\lambda , \quad
ed^{I_{1}} \rightarrow ed^{I_{1}}-4\lambda \;.
\label{10/13/2a}
\end{eqnarray}
To fix freedom of 
the conformal diffeomorphism invariance (\ref{10/13/2a}),
we set gauge fixing condition 
\begin{eqnarray}
h^{I_{1}\;\rho}_{\;\;\rho}=\alpha\;\pi^{I_{1}}\;, 
\label{10/13/3}
\end{eqnarray}
where $\alpha$ is gauge fixing parameter.
\footnote{More general gauge fixing condition; 
$h^{I_{1}\;\rho}_{\;\;\rho}=
\alpha\;\pi^{I_{1}}+\beta\nabla^2_{x}\pi^{I_{1}}
+\gamma\;ed^{I_{1}}+\varepsilon\nabla^2_{x}ed^{I_{1}}$
is also possible.}
The gauge symmetry (\ref{10/13/2a}) 
is not fixed completely
under gauge fixing condition (\ref{10/13/3}).
The residual gauge symmetry is 
generated by function $\lambda$ which satisfy
\begin{eqnarray}
\nabla^2_{x}\;\lambda+
\left(\frac{25}{3}-5\alpha\right)e^2\;\lambda
= 0 \;.
\label{10/13/4}
\end{eqnarray}
Substituting the gauge condition (\ref{10/13/3})
into (\ref{11/25/2}) and (\ref{11/25/3}), 
one can obtain
two scalar field equations.
After diagonalization of mass matrix,
one of the two equations becomes (\ref{10/13/1}).
The other equation becomes
\begin{eqnarray}
\nabla^2_{x}\;
\left(\pi^{I_{1}}-\frac{60}{8-3\alpha}ed^{I_{1}}\right)+
\left(\frac{25}{3}-5\alpha\right)e^2\;
\left(\pi^{I_{1}}-\frac{60}{8-3\alpha}ed^{I_{1}}\right)
= 0 \;.
\label{11/8/1}
\end{eqnarray}
It has the same form as (\ref{10/13/4}).
The solutions of (\ref{11/8/1}) 
are actually pure gauge modes!  

As a result,
setting the gauge fixing condition (\ref {10/13/3}),
we obtain two diagonalized equations 
(\ref {10/13/1}) and (\ref {11/8/1})
at $a=0$.
One finds physical modes with mass $m^2=45e^2$ 
in (\ref {10/13/1}) and 
pure gauge modes of conformal diffeomorphism
with mass 
$m^2=\left(\frac{25}{3}-5\alpha\right)e^2$
in (\ref {11/8/1}). 
The pure gauge modes can be gauged away 
and decouple from other physical modes.
\\

Next, we discuss connection between $a\neq 0$ and $a=0$.

Around $a\sim 0$, 
two scalar field equations 
(\ref{E3.1b}) and (\ref{E1.1b}) can be written as 
\begin{eqnarray}
\nabla^2_{x}\;\pi^{I_{1}}
+\left(\frac{5}{2}+\delta_{1}\right)\;\nabla^2_{x}\;ed^{I_{1}}
-e^2\left(45+\delta_{2}\right)\;\pi^{I_{1}}
-e^2\left(\frac{225}{2}+\delta_{3}\right)\;(ed^{I_{1}})
+\cdots=0,
\label{10/18/1} \\
\nabla^2_{x}\;\pi^{I_{1}}
+\left(\frac{5}{2}+\Delta_{1}\right)\;\nabla^2_{x}\;ed^{I_{1}}
-e^2\left(45+\Delta_{2}\right)\;\pi^{I_{1}}
-e^2\left(\frac{225}{2}+\Delta_{3}\right)\;(ed^{I_{1}})
+\cdots=0.
\label{10/18/1a} 
\end{eqnarray}
Here $''\cdots''$ means derivative and 
mixing terms with other tensors,
and $\delta, \Delta$ are $O(a^4u^4)$.
\footnote{
Readers should not confuse these $\Delta$'s
with Hodge-de Rham operator in Appendix A.
}
From (\ref {10/18/1}) and (\ref {10/18/1a}),
one can obtain
\begin{eqnarray}
\nabla^2_{x}\;(\pi^{I_{1}}+\frac{5}{2}(ed^{I_{1}}))
-45e^2(\pi^{I_{1}}+\frac{5}{2}(ed^{I_{1}}))
+\cdots = 0\;,
\label{10/18/3} \\
\nabla^2_{x}\;\left(\pi^{I_{1}}+
\frac{-45+\Delta_{13}}{\Delta_{12}}\;ed^{I_{1}}\right)
-e^2\left(\frac{5}{2}\Delta_{12}-\Delta_{13}\right)
\left(\pi^{I_{1}}+
\frac{-45+\Delta_{13}}{\Delta_{12}}\;ed^{I_{1}}\right)
+\cdots = 0 \;,
\label{10/18/3a}
\end{eqnarray}
where $\Delta_{12}=\frac{\delta_{2}-\Delta_{2}}{\delta_{1}-\Delta_{1}}=\frac{3}{16},
\Delta_{13}=\frac{\delta_{3}-\Delta_{3}}{\delta_{1}-\Delta_{1}}=-\frac{45}{4}$ .

Solutions of (\ref{10/18/3}) are physical modes with 
mass $m^2=45e^2$. 
On the other hand, one can find that 
solutions of (\ref{10/18/3a}) become gauge modes 
and can be gauged away
in the limit $a\rightarrow 0$. 
In this limit, 
any Kaluza-Klein modes of type IIB supergravity
compactified on $AdS_{5}\times S_{5}$ fall into
unitary irreducible representations of 
supergroup $U(2,2/4)$\cite{GM,GMZ}.
These decoupled scalars, which form {\bf 6} of $SO(6)$, 
are identified with scalars in the doubleton multiplet
in unitary irreducible representation of $U(2,2/4)$. 

This is just what we expect.
Doubleton scalars become pure gauge modes and decouple
in the limit $a\rightarrow 0$. \\

Finally, we mention the doubleton 2-form.

For $a=0$, 2-form field equations 
(\ref{(NSNS2.1)}) and (\ref{(RR2.1)}) 
in Appendix C
can be written in a factorized form \cite{KRN} : 
\begin{eqnarray}
(2ekI+i^{*}D)(2e(k+4)-i^{*}D)(b_{\mu\nu}+ia_{\mu\nu}) 
\quad =\quad  0
\label{11/8/2}
\end{eqnarray}
where ${}^{*}D\;a_{\mu\nu}=
\epsilon_{\mu\nu}^{\quad\rho\sigma\tau}
\nabla_{\rho}a_{\sigma\tau}$.
(\ref{11/8/2}) implies that 
$k=0$ mode is a pure gauge, and can be gauged away. 
This mode is the 2-form in the doubleton multiplet.

When $a\neq 0$, however, 
solutions of (\ref{11/8/2}) are not
the pure gauge mode only,
because the right hand side of (\ref{11/8/2}) is not zero.
So the decoupling of 2-form
dose not occurs.

\section{Summary and discussion}
In this paper, we consider 
the type IIB supergravity compactified 
on the gravity solution (\ref{5/31/1}) 
at the linearized level.

5d field equations (in Appendix C) become
complicated form with field mixing. 
The mixing (or mass) matrices 
obtained from the 5d field equations 
are functions of $u$ which is a 
coordinate on ${\cal M}_{5}$. 
This complication derives from lack of 
isometry, like conformal symmetry.
Unlike the $AdS$ space, 
the particle mass is not preserved by symmetry
in the gravity solution (\ref{5/31/1}),
due to the lack of the dilatation invariance.
This property corresponds to 
the fact that 
dimension of operators may be ambiguous 
in non-commutative Yang-Mills.

We find that the doubletons do not decouple
unless  $a=0$.
At $a=0$, doubletons couple with $U(1)$ operators 
of commutative Yang-Mills :
$\Phi^{I}, F_{\mu\nu},
\lambda_{\alpha}$ and $ \lambda^{\bar{\alpha}}$, 
on the boundary  
in gauge invariant way. 
But the gravity in the interior of boundary
has no information about $U(1)$ operators,
because doubletons decouple from other fields
in the interior of the boundary. 
On the other hand, 
if $a\neq 0$,
we can obtain the information 
of these $U(1)$ operators.
\footnote{Supergravity fields 
may also couple to operators of non-commutative theory, 
in a gauge invariant way.} 
Therefore, the gravity description is 
dual to $U(N)$ theory.
It is compatible with the fact that
$U(1)$ and $SU(N)$ gauge symmetries 
cannot be separated
in $U(N)$ non-commutative 
Yang-Mills theory. \\

Finally, we mention related questions. 

(i) 4d ${\cal N}=2$ super Yang-Mills theory 
was constructed 
by brane configuration \cite{Wit}.
In this construction, 
the gauge symmetry $U(1)$ of 
$U(N)\simeq U(1)\times SU(N)$ is ''frozen out''.
There is a question whether the frozen out occurs 
or not in presence of $B$-field.

(ii) Recently, non-commutative version of
$SU(N)$ Yang-Mills theory is proposed \cite{CD}.
The D-brane interpretation of this theory
may be interesting.    \\
\\

{\bf Acknowledgements}

I would to like thank Nobuyoshi Ohta and Kiyoshi Higashijima
for discussions and advice, 
and thank Tetsuji Kimura for careful reading 
for the manuscript.

\newpage
\section*{Appendix A.Notations}
\begin{itemize}
\item  $x^{A}$ with capital indices $A,B,...$ 
means 10d coordinates.
\item  $x^{\mu}$ with Greek indices $\mu,\nu,...$ 
means 5d coordinates 
$x^{0}, x^{1}, x^{2}, x^{3}$ and $u$.
\item  $y^{m}$ with Latin indices $m,n,...$ 
means coordinates of 5d sphere.
\end{itemize}

$\eta_{AB}=(-1,1,....,1)$ . \\

Covariant derivative : $\nabla_{A}$ . \\ 

The Riemann curvature tensor is defend by :
$[\nabla_{A},\nabla_{B}]\;V_{C}=
R_{AB\;\;C}^{\quad D}\;V_{D}$ 
with respect to vector $V_{A}$.
The Ricc tensor is defined by :
$R_{AB}=g^{CD}R_{ACBD}$.
The scalar curvature is defined by : 
$R=g^{AB}R_{AB}$.\\

Laplacian :
$\nabla_{x}^2\equiv g^{\rho\sigma}\nabla_{\rho}\nabla_{\sigma}\;,\quad
\nabla_{y}^2\equiv g^{mn}\nabla_{m}\nabla_{n}$\\

Spherical harmonics :
\begin{itemize}
\item $Y^{I_{1}}(y)$ is scalar.
\item $Y^{I_{5}}_{m}(y)$ is vector.
\item $Y^{I_{10}}_{[mn]}(y)$ is second rank 
anti-symmetric tensor.
\item $Y^{I_{14}}_{(mn)}(y)$ is second rank 
symmetric traceless tensor.
\end{itemize}
These spherical harmonics are eigenfunctions of
the Hodge-de Rham operator $\Delta$
and satisfy
\begin{eqnarray*}
&&\Delta\;Y^{I}=\nabla_{y}^2\;Y^{I}
=-e^2\;k(k+4)\;Y^{I} \quad k=0,1,...
\label{HRop1} \\
&&\Delta\;Y_{m}^{I_{5}}=(\nabla_{y}^2-4e^2)\;Y_{m}^{I_{5}}
=-e^2(k+1)(k+3)\;Y_{m}^{I_{5}} \quad k=1,2,....
\label{HRop2} \\
&&\Delta\;Y_{[mn]}^{I_{10}}=
(\nabla_{y}^2-6e^2)\;Y_{[mn]}^{I_{10}}
=-e^2(k+2)^2\;Y^{I_{10}}_{[mn]} \quad k=1,2,...
\label{HRop3} \\
&&\Delta\;Y_{(mn)}^{I_{14}}=
(\nabla_{y}^2-10e^2)\;Y_{(mn)}^{I_{14}}
=-e^2(k^2+4k+8)\;Y^{I} \quad k=2,3,... \;,
\label{HRop4} 
\end{eqnarray*}
where $e^2=1/\alpha^{\prime}R^2$.

\section*{Appendix B. Linearized field equations}
Linearized equations for small fluctuation around 
gravity solution (\ref{5/31/1}) are given in 
this section.\\

Definitions of 3- and 5-form field strengths are 
as follows :
\begin{eqnarray*}
{\cal H}_{\mu\nu\rho}\equiv 
3\;\partial_{[\mu}{\cal B}_{\nu\rho]}\;,\quad 
{\cal F}_{\mu\nu\rho}\equiv 
3\;\partial_{[\mu}{\cal A}_{\nu\rho]} \;,
\label{9/27/2}
\end{eqnarray*}

\begin{eqnarray*}
\tilde{{\cal F}}_{\mu\nu\rho\sigma\tau}&=&
5\;\partial_{[\mu}{\cal D}_{\nu\rho\sigma\tau]}
-\frac{1}{2}\;
\frac{5!}{2!3!}\;{\cal A}_{[\mu\nu}H_{\rho\sigma\tau]}
-\frac{1}{2}\;\frac{5!}{2!3!}\;A_{[\mu\nu}{\cal H}_{\rho\sigma\tau]} 
\nonumber \\
&&\qquad \qquad \qquad 
+\frac{1}{2}\;
\frac{5!}{2!3!}\;{\cal B}_{[\mu\nu}F_{\rho\sigma\tau]}
+\frac{1}{2}\;\frac{5!}{2!3!}\;B_{[\mu\nu}{\cal F}_{\rho\sigma\tau]} \;,
\label{9/27/3} \\
\nonumber \\
\tilde{{\cal F}}_{\mu\nu\rho\sigma m}&=&
\partial_{m}{\cal D}_{\mu\nu\rho\sigma}
+4\;\partial_{[\mu}{\cal D}_{\nu\rho\sigma] m}
\nonumber \\
&&-\frac{1}{2}\;\frac{4!}{3!}\;{\cal A}_{m[\mu}H_{\nu\rho\sigma]}
-\frac{1}{2}\;\frac{4!}{2!2!}\;A_{[\mu\nu}{\cal H}_{\rho\sigma] m} 
+\frac{1}{2}\;\frac{4!}{3!}\;{\cal B}_{m[\mu}F_{\nu\rho\sigma]}
+\frac{1}{2}\;\frac{4!}{2!2!}\;B_{[\mu\nu}{\cal F}_{\rho\sigma] m} \;,
\label{9/27/4} \\ 
\tilde{{\cal F}}_{\mu\nu\rho mn}&=&
2\;\partial_{[m}{\cal D}_{n]\mu\nu\rho}
+3\;\partial_{[\mu}{\cal D}_{\nu\rho]mn}
\nonumber \\
&&-\frac{1}{2}\;A_{mn}H_{\mu\nu\rho}
-\frac{1}{2}\;3\;A_{[\mu\nu}{\cal H}_{\rho]mn}
+\frac{1}{2}\;B_{mn}F_{\mu\nu\rho}
-\frac{1}{2}\;3\;B_{[\mu\nu}{\cal F}_{\rho]mn} \;,
\label{9/27/5} \\
\tilde{{\cal F}}_{\mu\nu mnp}&=&
3\;\partial_{[m}{\cal D}_{np]\mu\nu}
+2\;\partial_{[\mu}{\cal D}_{\nu]mnp}
-\frac{1}{2}\;A_{\mu\nu}{\cal H}_{mnp}
+\frac{1}{2}\;B_{\mu\nu}{\cal F}_{mnp} \;,
\label{9/25/6} \\
\tilde{{\cal F}}_{\mu mnpq}&=&
4\;\partial_{[m}{\cal D}_{npq]\mu}
+\partial_{\mu}{\cal D}_{mnpq}
\label{9/27/7} \\
\tilde{{\cal F}}_{mnpqr}&=&5\;\partial_{[m}{\cal D}_{npqr]}
\;.
\label{9/27/8} 
\end{eqnarray*}
\\
\begin{itemize}
\item Linearized scalar field equation from 
(\ref{3/18/1}) and (\ref{3/20/1}) : 
\end{itemize}
\begin{eqnarray}
\hspace{-30mm}
(\nabla_{x}^2+\nabla_{y}^2)\;\phi
-2\nabla\Phi\nabla\phi
+\left(\frac{11}{4}\nabla_{x}^2\Phi
-\frac{3}{2}(\nabla\Phi)^2
-\frac{13}{48}H^2-\frac{9}{48}e^{2\Phi}F^2+R
\right)\phi
+\frac{1}{6}(H{\cal H}-e^{2\Phi}F{\cal F})
\nonumber \\
+\left(-\frac{4}{3}\nabla_{x}^2\Phi
+\frac{2}{9}H^2+\frac{1}{18}e^{2\Phi}F^2
+\frac{1}{36}\tilde{F}_{\mu\nu\rho\tau}\tilde{F}^{\mu\nu\rho\tau}
-\frac{1}{60}\tilde{F}_{mnpqr}\tilde{F}^{mnpqr}
+\frac{2}{3}g^{\rho\sigma}R_{\rho\sigma}
+\frac{2}{5}g^{mn}R_{mn}
\right)h^{m}_{\;m}
\nonumber \\
+2\left(\frac{1}{4}\nabla_{x}^2\Phi
-\frac{1}{2}(\nabla\Phi)^2
+\frac{1}{48}(H^2+e^{2\Phi}F^2)
\right)h^{\prime\;\rho}_{\;\rho}
-\nabla^{\rho}\Phi(\nabla^{\sigma}h^{\prime}_{\rho\sigma}
-\frac{1}{2}\nabla_{\rho}h^{\prime\;\sigma}_{\sigma})
\nonumber \\
\hspace{-20mm}
+\left(3\nabla^{\rho}\nabla^{\sigma}\Phi
+2\nabla^{\rho}\Phi\nabla^{\sigma}\Phi
-\frac{3}{4}H_{\mu\nu}^{\;\;\rho}H^{\mu\nu\sigma}
-\frac{1}{4}e^{2\Phi}F_{\mu\nu}^{\;\;\rho}F^{\mu\nu\sigma}
-\frac{1}{12}\tilde{F}_{\mu\nu\lambda\tau}^{\qquad \rho}
\tilde{F}^{\mu\nu\lambda\tau\sigma}
\right)h^{\prime}_{\rho\sigma}
\quad =\quad 0\;,
\label{NS0} \\ \nonumber \\
(\nabla_{x}^2+\nabla_{y}^2 )\;c
-\frac{1}{3!}H^2 \;c
+\frac{1}{3!}(F{\cal H}+{\cal F}H)
\quad =\quad 0\;.
\qquad \qquad 
\label{R0}
\end{eqnarray}
\\

\begin{itemize}
\item Linearized Einstein equation from (\ref{3/18/5}) :
\end{itemize}
\begin{eqnarray}
&&
\frac{1}{2}(\nabla_{x}^2+\nabla_{y}^2)\;h^{\prime}_{\mu\nu}
+\frac{1}{2}\nabla_{\mu}\nabla_{\nu}h^{\prime\;\rho}_{\;\;\rho}
-\frac{1}{2}(\nabla_{\mu}\nabla^{\rho}h^{\prime}_{\rho\nu}
+\nabla_{\nu}\nabla^{\rho}h^{\prime}_{\rho\mu})
+R_{\mu\;\nu}^{\;\;\rho\;\;\sigma}\;h^{\prime}_{\rho\sigma}
-\frac{1}{2}(R_{\mu}^{\;\;\rho}h^{\prime}_{\nu\rho}
+R_{\nu}^{\;\;\rho}h^{\prime}_{\mu\rho})
\nonumber \\
&&-\left(\frac{1}{2}
H_{\mu\alpha}^{\quad\rho}H_{\nu}^{\;\;\alpha\sigma}
+\frac{1}{2}e^{2\Phi}
F_{\mu\alpha}^{\quad\sigma}F_{\nu}^{\;\;\alpha\sigma}
+\frac{1}{4!}e^{2\Phi}
\tilde{F}_{\mu\alpha\beta\gamma}^{\quad\;\;\;\rho}
F_{\nu}^{\;\;\alpha\beta\gamma\sigma}
\right)\;h^{\prime}_{\rho\sigma}
\nonumber \\
&&-\left(\frac{1}{4}\nabla_{x}^2\Phi
-\frac{1}{2}(\nabla\Phi)^2
+\frac{1}{48}H^2+\frac{1}{48}e^{2\Phi}F^{2}
\right)\;h^{\prime}_{\mu\nu}
-\frac{1}{4}g_{\mu\nu}\nabla^{\sigma}\Phi\;(
\nabla^{\rho}h^{\prime}_{\rho \sigma}
-\frac{1}{2}\nabla_{\sigma}h^{\prime\;\rho}_{\;\;\rho})
\nonumber \\
&&+g_{\mu\nu}\left(
\frac{1}{4}\nabla^{\rho}\nabla^{\sigma}\Phi
-\frac{1}{2}\nabla^{\rho}\Phi\nabla^{\sigma}\Phi
+\frac{1}{16}H_{\alpha\beta}^{\quad\rho}H^{\alpha\beta\sigma}
+\frac{1}{16}e^{2\Phi}F_{\alpha\beta}^{\quad\rho}F^{\alpha\beta\sigma}
\right)\;h^{\prime}_{\rho\sigma}
\nonumber \\
\nonumber \\
&&-\frac{1}{6}g_{\mu\nu}(\nabla_{x}^2+\nabla_{y}^2)\;
h^{m}_{\;m}
+\frac{1}{3}\left(
\frac{1}{2}H_{\mu\rho\sigma}H_{\nu}^{\;\;\rho\sigma}
+\frac{1}{2}e^{2\Phi}F_{\mu\rho\sigma}F_{\nu}^{\;\;\rho\sigma}
+\frac{1}{4!}e^{2\Phi}\tilde{F}_{\mu\rho\sigma\lambda\tau}
\tilde{F}_{\nu}^{\;\;\rho\sigma\lambda\tau}
\right)\;h^{m}_{\;m}
\nonumber \\
&&-\frac{1}{3}g_{\mu\nu}\;\left(
\frac{1}{4}\nabla_{x}^2\Phi-\frac{1}{2}(\nabla\Phi)^2
+\frac{1}{16}H^2+\frac{1}{16}e^{2\Phi}F^2
\right)\;h^{m}_{\;m}
\nonumber \\
\nonumber \\
&&+\frac{1}{2}g_{\mu\nu}
\nabla^{\rho}\Phi\nabla_{\rho}\phi
-\frac{1}{2}H_{\mu\rho\sigma}H_{\nu}^{\;\;\rho\sigma}\phi
+g_{\mu\nu}\left(\frac{1}{16}H^2
+\frac{1}{48}e^{2\Phi}F^2\right)\;\phi
\nonumber \\
\nonumber \\
&&+\frac{1}{4}({\cal H}_{\mu\rho\sigma}H_{\nu}^{\;\;\rho\sigma}
+H_{\mu\rho\sigma}{\cal H}_{\nu}^{\;\;\rho\sigma})
-\frac{1}{24}g_{\mu\nu}H{\cal H}
+\frac{1}{4}e^{2\Phi}({\cal F}_{\mu\rho\sigma}F_{\nu}^{\;\;\rho\sigma}
+F_{\mu\rho\sigma}{\cal F}_{\nu}^{\;\;\rho\sigma})
-\frac{1}{24}g_{\mu\nu}e^{2\Phi}F{\cal F}
\nonumber \\
\nonumber \\
&&+\frac{1}{4\cdot4!}e^{2\Phi}(
\tilde{{\cal F}}_{\mu\rho\sigma\lambda\tau}
\tilde{F}_{\nu}^{\;\;\rho\sigma\lambda\tau}
+\tilde{F}_{\mu\rho\sigma\lambda\tau}
\tilde{{\cal F}}_{\nu}^{\;\;\rho\sigma\lambda\tau})
\quad =\quad 0\;,
\qquad\qquad \label{Ein1} 
\end{eqnarray}

\begin{eqnarray}
&&\frac{1}{2}(\nabla_{x}^2+\nabla_{y}^2)\;h_{\mu m}
-\frac{1}{2}\nabla_{\mu}\nabla^{\rho}h_{\rho m}
-\frac{1}{2}\nabla_{m}\nabla^{n}h_{n\mu}
-\frac{1}{2}R_{\mu}^{\;\;\rho}h_{\rho m}
-\frac{1}{2}R_{m}^{\;\;n}h_{n\mu}
\nonumber \\
&&
-\left(\frac{1}{4}\nabla_{x}^2\Phi
-\frac{1}{2}(\nabla\Phi)^2
+\frac{1}{48}H^{\nu\rho\sigma}H_{\nu\rho\sigma}
+\frac{1}{48}e^{2\Phi}F^{\nu\rho\sigma}F_{\nu\rho\sigma}
\right)\;h_{\mu m}
\nonumber \\
&&-\frac{1}{2}\nabla_{m}\nabla^{\rho}h^{\prime}_{\rho\mu}
+\frac{1}{2}\nabla_{\mu}\nabla_{m}h^{\prime\;\rho}_{\;\;\rho}
-\frac{4}{15}\nabla_{\mu}\nabla_{m}h^{m}_{\;\;m}
-\frac{1}{2}\nabla_{\mu}\nabla^{n}h_{(nm)}
\nonumber \\
&&
+\frac{1}{4}H_{\mu}^{\;\rho\sigma}
{\cal H}_{m\rho\sigma}
+\frac{1}{4}e^{2\Phi}F_{\mu}^{\;\rho\sigma}
{\cal F}_{m\rho\sigma}
\nonumber \\
&&+\frac{1}{4\cdot4!}e^{2\Phi}\tilde{F}_{\mu}^{\;\;\nu\rho\sigma\lambda}
\tilde{{\cal F}}_{m\nu\rho\sigma\lambda}
+\frac{1}{4\cdot4!}e^{2\Phi}\tilde{F}_{m}^{\;\;nlkj}
\tilde{{\cal F}}_{\mu nlkj}
\quad =\quad 0\;,
\qquad\qquad \label{Ein2} 
\end{eqnarray}

\begin{eqnarray}
&&\frac{1}{2}(\nabla_{x}^2+\nabla_{y}^2)\;h_{(mn)}
-\frac{1}{2}\nabla_{m}\nabla^{l}h_{(ln)}
-\frac{1}{2}\nabla_{n}\nabla^{l}h_{(lm)}
+R_{m\;n}^{\;\;l\;\;k}\;h_{(lk)}
-\frac{1}{2}R_{m}^{\;\;l}\;h_{(ln)}
-\frac{1}{2}R_{n}^{\;\;l}\;h_{(lm)}
\nonumber \\
&&-\frac{1}{4!}\tilde{F}_{mlkj}^{\quad\;\;\; a}
\tilde{F}_{n}^{\;\;lkjb}
\;h_{(ab)}
-\left(\frac{1}{4}\nabla_{x}^2\Phi
-\frac{1}{2}(\nabla\Phi)^2
+\frac{1}{48}H^{\mu\nu\rho}H_{\mu\nu\rho}
+\frac{1}{48}e^{2\Phi}F^{\mu\nu\rho}F_{\mu\nu\rho}
\right)\;h_{(mn)}
\nonumber \\
&&+g_{mn}
\left(\frac{1}{4}\nabla^{\mu}\nabla^{\nu}\Phi
-\frac{1}{2}\nabla^{\mu}\Phi\nabla^{\nu}\Phi
+\frac{1}{16}H^{\mu}_{\;\;\rho\sigma}H^{\nu\rho\sigma}
+\frac{1}{16}e^{2\Phi}F^{\mu}_{\;\;\rho\sigma}F^{\nu\rho\sigma}
\right)
\;h^{\prime}_{\mu\nu}
+\frac{1}{2}\nabla_{m}\nabla_{n}h^{\prime\;\rho}_{\;\;\rho}
\nonumber \\
&&+\frac{1}{10}g_{mn}(\nabla_{x}^2+\nabla_{y}^2)\;h^{l}_{\;\;l}
-\frac{8}{15}\nabla_{m}\nabla_{n}h^{l}_{\;\;l}
-\frac{1}{5!}e^{2\Phi}\tilde{F}_{m}^{\;abcd}\tilde{F}_{nabcd}\;h^{l}_{\;\;l}
\nonumber \\
&&-g_{mn}\left(\frac{2}{15}\nabla_{x}^2\Phi
-\frac{4}{15}(\nabla\Phi)^2
+\frac{1}{40}H^{\mu\nu\rho}H_{\mu\nu\rho}
+\frac{1}{40}e^{2\Phi}F^{\mu\nu\rho}F_{\mu\nu\rho}
\right)\;h^{l}_{\;\;l}
\nonumber \\
&&-\frac{1}{2}\nabla_{m}\nabla^{\rho}h_{\rho n}
-\frac{1}{2}\nabla_{n}\nabla^{\rho}h_{\rho m}
\nonumber \\
&&+\frac{1}{2}g_{mn}\nabla^{\mu}\Phi\nabla_{\mu}\phi
+g_{mn}
\left(\frac{1}{16}H^{\mu\nu\rho}H_{\mu\nu\rho}
+\frac{1}{48}e^{2\Phi}F^{\mu\nu\rho}F_{\mu\nu\rho}
\right)\;\phi
\nonumber \\
&&-\frac{1}{24}g_{mn}
(H^{\mu\nu\rho}{\cal H}_{\mu\nu\rho}
+e^{\Phi}F^{\mu\nu\rho}{\cal F}_{\mu\nu\rho})
+\frac{1}{4!}e^{2\Phi}\tilde{F}_{m}^{\;abcd}\tilde{{\cal F}}_{nabcd}
+\frac{1}{4!}e^{2\Phi}\tilde{F}_{n}^{\;abcd}\tilde{{\cal F}}_{mabcd}
\quad =\; 0.
\quad \label{Ein3} 
\end{eqnarray}
\\
  
\begin{itemize}
\item Linearized NS-NS 2-form equation from (\ref{3/18/4}) :
\end{itemize}
\begin{eqnarray}
&&\nabla^{\rho}{\cal H}_{\rho\mu\nu}
+\nabla_{m}{\cal H}_{m\mu\nu}
-2\nabla^{\rho}\Phi{\cal H}_{\rho\mu\nu}
+H_{\mu\nu\rho}\nabla^{\rho}c
+\nabla^{\rho}H_{\rho\mu\nu}\;c
\nonumber \\
&&
-\frac{9}{4}\nabla^{\rho}\phi\;H_{\rho\mu\nu}
-\left(\frac{5}{2}\nabla^{\rho}H_{\rho\mu\nu}
-5\nabla^{\rho}\Phi H_{\rho\mu\nu}
+\frac{1}{4}\tilde{F}_{\mu\nu\rho\sigma\tau}
F^{\rho\sigma\tau}
\right)\phi
\nonumber \\
&&+\frac{2}{3}H_{\mu\nu\rho}\nabla^{\rho}h^{m}_{\;m}
+\left(\frac{1}{3}\nabla^{\rho}H_{\rho\mu\nu}
-\frac{2}{3}\nabla^{\rho}\Phi H_{\rho\mu\nu}
+\frac{1}{6}e^{2\Phi}\tilde{F}_{\mu\nu\rho\sigma\tau}
F^{\rho\sigma\tau}\right)h^{m}_{\;m}
\nonumber \\
&&
-H_{\sigma\mu\nu}(\nabla^{\rho}h^{\prime\;\sigma}_{\;\rho}
-\frac{1}{2}\nabla^{\sigma}h^{\prime\;\rho}_{\;\rho})
-H_{\rho\sigma\nu}\nabla^{\rho}h^{\prime\;\sigma}_{\;\mu}
-H_{\rho\mu\sigma}\nabla^{\rho}h^{\prime\;\sigma}_{\;\nu}
\nonumber \\
&&
+\left(-\nabla^{\rho}H^{\sigma}_{\;\mu\nu}
+2\nabla^{\rho}\Phi H^{\sigma}_{\;\mu\nu}
-\frac{1}{2}e^{2\Phi}
\tilde{F}_{\mu\nu\lambda\tau}^{\quad\;\;\rho}
F^{\lambda\tau\sigma}
\right)h^{\prime}_{\rho\sigma}
\!\!
+\frac{1}{3!}e^{2\Phi}\tilde{F}_{\mu\nu}^{\quad\rho\sigma\tau}
{\cal F}_{\rho\sigma\tau}
+\frac{1}{3!}e^{2\Phi}F^{\rho\sigma\tau}
\tilde{{\cal F}}_{\mu\nu\rho\sigma\tau}
\nonumber \\
&&\quad =\quad 0 \,
\label{NS2.1}
\end{eqnarray}

\begin{eqnarray}
&&\nabla^{\rho}{\cal H}_{\rho\mu m}
+\nabla^{l}{\cal H}_{l\mu m}
-2\nabla^{\rho}\Phi\;{\cal H}_{\rho\mu m}
-\nabla_{\rho}h_{\sigma m}
H_{\mu}^{\;\rho\sigma}
+\frac{1}{3!}e^{2\Phi}\tilde{{\cal F}}_{\mu m\rho\sigma\lambda}
F^{\rho\sigma\lambda}
= 0\;,
\label{NS2.2} 
\end{eqnarray}
\begin{eqnarray}
&&\nabla^{\rho}{\cal H}_{\rho mn}
+\nabla^{l}{\cal H}_{lmn}
-2\nabla^{\rho}\Phi\;{\cal H}_{\rho mn}
+\frac{1}{3!}e^{2\Phi}\tilde{F}_{mnlkj}
{\cal F}^{lkj}
+\frac{1}{3!}e^{2\Phi}\tilde{{\cal F}}_{mn\rho\sigma\lambda}
F^{\rho\sigma\lambda}
=0.
\label{NS2.3} 
\end{eqnarray}
\\

\begin{itemize}
\item Linearized R-R 2-form equation from (\ref{3/18/3}) :
\end{itemize}
\begin{eqnarray}
&&
\nabla^{\rho}{\cal F}_{\rho\mu\nu}
+\nabla^{m}{\cal F}_{m\mu\nu}
-\frac{1}{4}F_{\rho\mu\nu}\nabla^{\rho}\phi
-\left(\frac{1}{2}\nabla^{\rho}F_{\rho\mu\nu}
+\frac{1}{4}\tilde{F}_{\mu\nu\rho\sigma\tau}
H^{\rho\sigma\tau}
\right)\phi
\nonumber \\
&&+H_{\mu\nu\rho}\nabla^{\rho}c
+\nabla^{\rho}H_{\rho\mu\nu}\;c
+\frac{2}{3}H_{\mu\nu\rho}\nabla^{\rho}h^{m}_{\;m}
+\left(\frac{1}{3}\nabla^{\rho}F_{\rho\mu\nu}
-\frac{1}{6}\tilde{F}_{\mu\nu\rho\sigma\tau}
H^{\rho\sigma\tau}\right)h^{m}_{\;m}
\nonumber \\
&&-F_{\sigma\mu\nu}(\nabla^{\rho}h^{I_{1}\;\sigma}_{\;\rho}
-\frac{1}{2}\nabla^{\sigma}h^{\prime\;\rho}_{\;\rho})
-F_{\rho\sigma\nu}\nabla^{\rho}h^{\prime\;\sigma}_{\;\mu}
-F_{\rho\mu\sigma}\nabla^{\rho}h^{\prime\;\sigma}_{\;\nu}
-\left(\nabla^{\rho}F^{\sigma}_{\;\mu\nu}
+\frac{1}{2}\tilde{F}_{\mu\nu\lambda\tau}^{\quad\;\;\rho}
H^{\lambda\tau\sigma}
\right)h^{\prime}_{\rho\sigma}
\nonumber \\
&&-\frac{1}{3!}\tilde{F}_{\mu\nu}^{\quad\rho\sigma\tau}
{\cal H}_{\rho\sigma\tau}
-\frac{1}{3!}H^{\rho\sigma\tau}
\tilde{{\cal F}}_{\mu\nu\rho\sigma\tau}
\quad =\quad 0\;,
\label{R2.1}
\end{eqnarray}

\begin{eqnarray}
&&\nabla^{\rho}{\cal F}_{\rho\mu m}
+\nabla^{l}{\cal F}_{l\mu m}
+2\nabla^{\rho}\Phi\;{\cal F}_{\rho\mu m}
-\nabla_{\rho}h_{\sigma m}\;F_{\mu}^{\;\rho\sigma}
-\frac{1}{3!}\tilde{{\cal F}}_{\mu m\rho\sigma\lambda}
H^{\rho\sigma\lambda}
\quad =\quad 0,
\label{R2.2} 
\end{eqnarray}
\begin{eqnarray}
&&\nabla^{\rho}{\cal F}_{\rho mn}
+\nabla^{l}{\cal F}_{lmn}
+2\nabla^{\rho}\Phi\;{\cal F}_{\rho mn}
-\frac{1}{3!}\tilde{F}_{mnlkj}
{\cal H}^{lkj}
-\frac{1}{3!}\tilde{{\cal F}}_{mn\rho\sigma\lambda}
H^{\rho\sigma\lambda}
\quad =\quad 0.
\label{R2.3} 
\end{eqnarray}
\\

\begin{itemize}
\item Linearized self-dual equation from (\ref{5/31/2}) : 
\end{itemize}
\begin{eqnarray}
\tilde{{\cal F}}_{\mu\nu\rho\sigma\lambda}&=&
\frac{1}{5!}\epsilon_{\mu\nu\rho\sigma\lambda}^{\qquad mnlkj}
\left[
\tilde{{\cal F}}_{mnlkj}
+\frac{1}{2}\tilde{F}_{mnlkj}\;h^{\prime\;\rho}_{\rho}
-\frac{4}{3}\tilde{F}_{mnlkj}\;h^{m}_{\;\;m}
-5\tilde{F}^{i}_{\;nlkj}\;h_{(im)}
\right]
\label{SD1}\;, \\
\tilde{{\cal F}}_{\mu\nu\rho\sigma m}&=&
\frac{5}{5!}\epsilon_{\mu\nu\rho\sigma m}^{\qquad\;\;\;nlkj\lambda}
\left[
\tilde{{\cal F}}_{nlkj\lambda}
-5\tilde{F}_{nlkj}^{\quad\; i}\;h_{i\lambda}
\right]
\label{SD2} \;, \\
\tilde{{\cal F}}_{\mu\nu\rho mn}&=&
\frac{10}{5!}\epsilon_{\mu\nu\rho mn}^{\qquad\;\;\;lkj\sigma\lambda}
\tilde{{\cal F}}_{lkj\sigma\lambda} \;.
\label{SD3} 
\end{eqnarray}
One obtains five linearized equations from (\ref{5/31/2}). 
Since two of the five equations are equivalent 
to other three, we write 
only three independent equations here.

\section*{Appendix C. Linearized equations expanded by Spherical harmonics}
Results of harmonic expansion are 
given in this section.\\

Definitions of differential operator $Max$ 
are as follows :
\begin{eqnarray*}
Max\;a^{I_{1}}_{\mu\nu}&\equiv& \nabla^{\rho}\;
3\;\partial_{[\rho}a^{I_{5}}_{\mu\nu]}=
\nabla^{\rho}(\nabla_{\rho}a^{I_{1}}_{\mu\nu}
+\nabla_{\mu}a^{I_{1}}_{\nu\rho}
+\nabla_{\nu}a^{I_{1}}_{\rho\mu}) \;,
\label{9/28/2} \\
\nonumber \\
Max\;a^{I_{5}}_{\mu}&\equiv& \nabla^{\rho}\;
2\;\partial_{[\rho}a^{I_{5}}_{\mu]}=
\nabla^{\rho}(\nabla_{\rho}a^{I_{5}}_{\mu}-\nabla_{\mu}a^{I_{5}}_{\rho}) 
\;.
\label{9/28/3}
\end{eqnarray*}

Definitions of 4-form $\tilde{D}^{I}$ and 5-form 
$\tilde{{\cal F}}^{I}$ are as follows
\begin{eqnarray*}
\tilde{D}^{I_{1}}_{\mu\nu\rho\sigma}&=&
d^{I_{1}}_{\mu\nu\rho\sigma}
-3\;A_{[\mu\nu}b^{I_{1}}_{\rho\sigma]}
+3\;B_{[\mu\nu}a^{I_{1}}_{\rho\sigma]}
\label{9/28/4} \\
\tilde{D}^{I_{5}}_{\mu\nu\rho}&=&
d^{I_{5}}_{\mu\nu\rho}
+\frac{3}{2}\;A_{[\mu\nu}b^{I_{5}}_{\rho]}
-\frac{3}{2}\;B_{[\mu\nu}a^{I_{5}}_{\rho]}
\label{9/28/5} \\
\tilde{D}^{I_{10}}_{\mu\nu}&=&
d^{I_{10}}_{\mu\nu}
-\frac{1}{2}A_{\mu\nu}b^{I_{10}}
+\frac{1}{2}B_{\mu\nu}a^{I_{10}} \;,
\label{9/28/6} \\
\nonumber \\
\tilde{{\cal F}}^{I_{1}}_{\mu\nu\rho\sigma\tau}&=&
5\;\partial_{[\;\mu}d^{I_{1}}_{\nu\rho\sigma\tau]}
-5\;a^{I_{1}}_{[\mu\nu}H_{\rho\sigma\tau]}
-5\;A_{[\mu\nu}\;
3\partial_{\rho}b^{I_{1}}_{\sigma\tau]}
+5\;b^{I_{1}}_{[\mu\nu}F_{\rho\sigma\tau]}
+5\;B_{[\mu\nu}\;
3\partial_{\rho}a^{I_{1}}_{\sigma\tau]}
\label{9/28/7} \\
\tilde{{\cal F}}^{I_{5}}_{\mu\nu\rho\sigma}&=&
4\;\partial_{[\mu}d^{I_{5}}_{\nu\rho\sigma]}
+2\;a^{I_{5}}_{[\mu}H_{\nu\rho\sigma]}
-3\;A_{[\mu\nu}\;2\partial_{\rho}b^{I_{5}}_{\sigma]}
-2\;b^{I_{5}}_{[\mu}F_{\nu\rho\sigma]}
+3\;B_{[\mu\nu}\;2\partial_{\rho}a^{I_{5}}_{\sigma]}
\label{9/28/8} \\
\tilde{{\cal F}}^{I_{10}}_{\mu\nu\rho}&=&
3\;\partial_{[\mu}d^{I_{10}}_{\nu\rho]}
-\frac{1}{2}\;H_{\mu\nu\rho}a^{I_{10}}
-\frac{3}{2}\;A_{[\mu\nu}\partial_{\rho]}b^{I_{10}}
+\frac{1}{2}\;F_{\mu\nu\rho}b^{I_{10}}
+\frac{3}{2}\;B_{[\mu\nu}\partial_{\rho]}a^{I_{10}}
\;, \label{9/28/9} 
\end{eqnarray*}
\\

\begin{itemize}
\item Harmonic expansions of 
linearized scalar field equations 
(\ref{NS0}) and (\ref{R0}) :
\end{itemize}
\begin{eqnarray}
\hspace{-30mm}
\Bigg\{(\nabla_{x}^2+\nabla_{y}^2)\;\phi^{I_{1}}
-2\nabla\Phi\nabla\phi^{I_{1}}
+\left(\frac{11}{4}\nabla_{x}^2\Phi
-\frac{3}{2}(\nabla\Phi)^2
-\frac{13}{48}H^2-\frac{9}{48}e^{2\Phi}F^2+R
\right)\phi^{I_{1}}
+\frac{1}{6}(H{\cal H}^{I_{1}}-e^{2\Phi}F{\cal F}^{I_{1}})
\nonumber \\
+\left(-\frac{4}{3}\nabla_{x}^2\Phi
+\frac{2}{9}H^2+\frac{1}{18}e^{2\Phi}F^2
+\frac{1}{36}\tilde{F}_{\mu\nu\rho\tau}\tilde{F}^{\mu\nu\rho\tau}
-\frac{1}{60}\tilde{F}_{mnpqr}\tilde{F}^{mnpqr}
+\frac{2}{3}g^{\rho\sigma}R_{\rho\sigma}
+\frac{2}{5}g^{mn}R_{mn}
\right)\pi^{I_{1}}
\nonumber \\
+2\left(\frac{1}{4}\nabla_{x}^2\Phi
-\frac{1}{2}(\nabla\Phi)^2
+\frac{1}{48}(H^2+e^{2\Phi}F^2)
\right)h^{I_{1}\;\rho}_{\;\rho}
-\nabla^{\rho}\Phi(\nabla^{\sigma}h^{I_{1}}_{\rho\sigma}
-\frac{1}{2}\nabla_{\rho}h^{I_{1}\;\sigma}_{\sigma})
\nonumber \\
\hspace{-20mm}
+\left(3\nabla^{\rho}\nabla^{\sigma}\Phi
+2\nabla^{\rho}\Phi\nabla^{\sigma}\Phi
-\frac{3}{4}H_{\mu\nu}^{\;\;\rho}H^{\mu\nu\sigma}
-\frac{1}{4}e^{2\Phi}F_{\mu\nu}^{\;\;\rho}F^{\mu\nu\sigma}
-\frac{1}{12}\tilde{F}_{\mu\nu\lambda\tau}^{\qquad \rho}
\tilde{F}^{\mu\nu\lambda\tau\sigma}
\right)h^{I_{1}}_{\rho\sigma}
\Bigg\}\;Y^{I_{1}}
\;=\;0 \;,
\label{NSNS0} \\ \nonumber \\
\Bigg\{(\nabla_{x}^2+\nabla_{y}^2 )\;c^{I_{1}}
-\frac{1}{3!}H^2 \;c^{I_{1}}
+\frac{1}{3!}(F{\cal H}^{I_{1}}+{\cal F}^{I_{1}}H)
\Bigg\}\;Y^{I_{1}}
\quad =\quad 0\;.
\qquad \qquad 
\label{RR0}
\end{eqnarray}
\\

\begin{itemize}
\item Harmonic expansions of linearized Einstein equations 
(\ref{Ein1}), (\ref{Ein2}) and (\ref{Ein3}) :
\end{itemize}
\begin{eqnarray}
&&\hspace{-10mm}\Bigg\{
\frac{1}{2}(\nabla_{x}^2+\nabla_{y}^2)\;h^{I_{1}}_{\mu\nu}
+\frac{1}{2}\nabla_{\mu}\nabla_{\nu}h^{I_{1}\;\rho}_{\;\;\rho}
-\frac{1}{2}(\nabla_{\mu}\nabla^{\rho}h^{I_{1}}_{\rho\nu}
+\nabla_{\nu}\nabla^{\rho}h^{I_{1}}_{\rho\mu})
+R_{\mu\;\nu}^{\;\;\rho\;\;\sigma}\;h^{I_{1}}_{\rho\sigma}
-\frac{1}{2}(R_{\mu}^{\;\;\rho}h^{I_{1}}_{\nu\rho}
+R_{\nu}^{\;\;\rho}h^{I_{1}}_{\mu\rho})
\nonumber \\
&&-\left(\frac{1}{2}
H_{\mu\alpha}^{\quad\rho}H_{\nu}^{\;\;\alpha\sigma}
+\frac{1}{2}e^{2\Phi}
F_{\mu\alpha}^{\quad\sigma}F_{\nu}^{\;\;\alpha\sigma}
+\frac{1}{4!}e^{2\Phi}
\tilde{F}_{\mu\alpha\beta\gamma}^{\quad\;\;\;\rho}
F_{\nu}^{\;\;\alpha\beta\gamma\sigma}
\right)\;h^{I_{1}}_{\rho\sigma}
\nonumber \\
&&-\left(\frac{1}{4}\nabla_{x}^2\Phi
-\frac{1}{2}(\nabla\Phi)^2
+\frac{1}{48}H^2+\frac{1}{48}e^{2\Phi}F^{2}
\right)\;h^{I_{1}}_{\mu\nu}
-\frac{1}{4}g_{\mu\nu}\nabla^{\sigma}\Phi\;(
\nabla^{\rho}h^{\prime}_{\rho \sigma}
-\frac{1}{2}\nabla_{\sigma}h^{I_{1}\;\rho}_{\;\;\rho})
\nonumber \\
&&+g_{\mu\nu}\left(
\frac{1}{4}\nabla^{\rho}\nabla^{\sigma}\Phi
-\frac{1}{2}\nabla^{\rho}\Phi\nabla^{\sigma}\Phi
+\frac{1}{16}H_{\alpha\beta}^{\quad\rho}H^{\alpha\beta\sigma}
+\frac{1}{16}e^{2\Phi}F_{\alpha\beta}^{\quad\rho}F^{\alpha\beta\sigma}
\right)\;h^{I_{1}}_{\rho\sigma}
\nonumber \\
\nonumber \\
&&-\frac{1}{6}g_{\mu\nu}(\nabla_{x}^2+\nabla_{y}^2)\;
\pi^{I_{1}}
+\frac{1}{3}\left(
\frac{1}{2}H_{\mu\rho\sigma}H_{\nu}^{\;\;\rho\sigma}
+\frac{1}{2}e^{2\Phi}F_{\mu\rho\sigma}F_{\nu}^{\;\;\rho\sigma}
+\frac{1}{4!}e^{2\Phi}\tilde{F}_{\mu\rho\sigma\lambda\tau}
\tilde{F}_{\nu}^{\;\;\rho\sigma\lambda\tau}
\right)\;\pi^{I_{1}}
\nonumber \\
&&-\frac{1}{3}g_{\mu\nu}\;\left(
\frac{1}{4}\nabla_{x}^2\Phi-\frac{1}{2}(\nabla\Phi)^2
+\frac{1}{16}H^2+\frac{1}{16}e^{2\Phi}F^2
\right)\;\pi^{I_{1}}
\nonumber \\
\nonumber \\
&&+\frac{1}{2}g_{\mu\nu}
\nabla^{\rho}\Phi\nabla_{\rho}\phi^{I_{1}}
-\frac{1}{2}H_{\mu\rho\sigma}H_{\nu}^{\;\;\rho\sigma}
\phi^{I_{1}}
+g_{\mu\nu}\left(\frac{1}{16}H^2
+\frac{1}{48}e^{2\Phi}F^2\right)\;\phi^{I_{1}}
\nonumber \\
\nonumber \\
&&+\frac{1}{4}({\cal H}^{I_{1}}_{\mu\rho\sigma}H_{\nu}^{\;\;\rho\sigma}
+H_{\mu\rho\sigma}{\cal H}_{\nu}^{I_{1}\;\;\rho\sigma})
-\frac{1}{24}g_{\mu\nu}H{\cal H}^{I_{1}}
\nonumber \\
&&+\frac{1}{4}e^{2\Phi}({\cal F}^{I_{1}}_{\mu\rho\sigma}F_{\nu}^{\;\;\rho\sigma}
+F_{\mu\rho\sigma}{\cal F}_{\nu}^{I_{1}\;\;\rho\sigma})
-\frac{1}{24}g_{\mu\nu}e^{2\Phi}F{\cal F}^{I_{1}}
\nonumber \\
\nonumber \\
&&+\frac{1}{4\cdot4!}e^{2\Phi}(
\tilde{{\cal F}}^{I_{1}}_{\mu\rho\sigma\lambda\tau}
\tilde{F}_{\nu}^{\;\;\rho\sigma\lambda\tau}
+\tilde{F}_{\mu\rho\sigma\lambda\tau}
\tilde{{\cal F}}_{\nu}^{I_{1}\;\;\rho\sigma\lambda\tau})
\Bigg\}\;Y^{I_{1}}
\quad =\quad 0\;, 
\label{E1} 
\end{eqnarray}

\begin{eqnarray}
&\Bigg\{&\!\!\!\!
\frac{1}{2}\nabla_{\mu}h^{I_{1}\;\;\nu}_{\nu}
-\frac{1}{2}\nabla^{\nu}h_{\nu \mu}^{I_{1}}
-\frac{4}{15}\;\pi^{I_{1}}
+\frac{1}{4}(H_{\mu}^{\;\;\rho\sigma}b_{\rho\sigma}^{I_{1}}
+e^{2\Phi}F_{\mu}^{\;\;\rho\sigma}a_{\rho\sigma}^{I_{1}})
+\frac{1}{4\cdot4!}\;
e^{2\Phi}\tilde{F}_{\mu}^{\;\;\nu\rho\sigma\tau}
\;\tilde{D}^{I_{1}}_{\nu\rho\sigma\tau}
\!\Bigg\}\nabla_{m}Y^{I_{1}}
\nonumber \\
&&+\frac{1}{4\cdot4!}\;e^{2\Phi}\tilde{F}_{m}^{\;\;npqr}\;
\nabla_{\mu}d^{I_{1}}
\;\epsilon_{npqr}^{\quad \;\;a}\;\nabla_{a}Y^{I_{1}}
\quad=\quad 0 \;, 
\label{E2.1} 
\end{eqnarray}

\begin{eqnarray}
&\Bigg\{&
\frac{1}{2}(Max+\nabla_{y}^2)\;h^{I_{5}}_{\mu}
-\left(\frac{1}{4}\nabla_{x}^2\Phi
-\frac{1}{2}(\nabla\Phi)^2
+\frac{1}{48}H^{\rho\sigma\lambda}H_{\rho\sigma\lambda}
+\frac{1}{48}e^{2\Phi}F^{\rho\sigma\lambda}F_{\rho\sigma\lambda}\right)
h_{\mu}^{I_{5}}
\nonumber \\
&&+\frac{1}{4}H_{\mu}^{\;\;\rho\sigma}\;
(\partial_{\rho}b_{\sigma}^{I_{5}}-\partial_{\sigma}b_{\rho}^{I_{5}})
+\frac{1}{4}e^{2\Phi}F_{\mu}^{\;\;\rho\sigma}\;
(\partial_{\rho}a_{\sigma}^{I_{5}}-\partial_{\sigma}a_{\rho}^{I_{5}})
+\frac{1}{4\cdot4!}\;e^{2\Phi}\tilde{F}_{\mu}^{\;\;\nu\rho\sigma\lambda}
\tilde{{\cal F}}^{I_{5}}_{\nu\rho\sigma\lambda}
\Bigg\}\;Y_{m}^{I_{5}}
\nonumber \\
&&+\frac{1}{4\cdot4!}\;e^{2\Phi}
\tilde{F}_{m}^{\;\;npqr}\;
d_{\mu}^{I_{5}}\;
\epsilon_{[pqr}^{\quad ab}\;\nabla_{n]}\nabla_{a}Y_{b}^{I_{5}}
\quad =\quad 0\;, 
\label{E2.2}  
\end{eqnarray}
\begin{eqnarray}
&&\hspace{-15mm}\Bigg\{\frac{1}{4}g_{mn}\nabla^{\rho}\Phi\;
(\nabla^{\sigma}h^{I_{1}}_{\sigma\rho}
-\frac{1}{2}\nabla_{\rho}h^{I_{1}\;\sigma}_{\;\;\sigma})
+g_{mn}\left(\frac{1}{4}\nabla^{\rho}\nabla^{\sigma}\Phi
-\frac{1}{2}\nabla^{\rho}\Phi\nabla^{\sigma}\Phi
+\frac{1}{16}(H_{\mu\nu}^{\;\;\rho}H^{\mu\nu\sigma}
+e^{2\Phi}F_{\mu\nu}^{\;\;\rho}F^{\mu\nu\sigma})
\right)\;h^{I_{1}}_{\rho\sigma}
\nonumber \\
&&+\frac{1}{10}g_{mn}(\nabla_{x}^2+\nabla_{y}^2)\;\pi^{I_{1}}
-\frac{8}{75}g_{mn}\nabla_{y}^2\;\pi^{I_{1}}
+\frac{1}{10}g_{mn}\nabla_{y}^2\;h^{I_{1}\;\rho}_{\;\;\rho}
\nonumber \\
&&-\frac{1}{5!}e^{2\Phi}\tilde{F}_{mpqrs}\tilde{F}_{n}^{\;\;pqrs}\;\pi^{I_{1}}
-g_{mn}\left(\frac{2}{15}\nabla_{x}^2\Phi
-\frac{4}{15}(\nabla\Phi)^2
+\frac{1}{40}(H^2+e^{2\Phi}F^2)\;
\right)\;\pi^{I_{1}}
\nonumber \\ \nonumber \\
&&+\frac{1}{2}g_{mn}
\nabla^{\mu}\Phi\nabla_{\mu}\phi^{I_{1}}
+g_{mn}\left(\frac{1}{16}H^2+\frac{1}{48}e^{2\Phi}F^2
\right)\phi^{I_{1}}
\nonumber \\
&&
-\frac{1}{24}g_{mn}(H^{\mu\nu\rho}{\cal H}^{I_{1}}_{\mu\nu\rho}
+e^{2\Phi}F^{\mu\nu\rho}{\cal F}^{I_{1}}_{\mu\nu\rho})
\qquad \Bigg\}\;Y^{I_{1}}
\nonumber \\
&&+\frac{5}{4\cdot4!}\tilde{F}_{n}^{\;\;pqrs}\;
d^{I_{1}}\nabla_{[m}\epsilon_{pqrs]}^{\quad\;\;a}\;
\nabla_{a}Y^{I_{1}}
+\frac{5}{4\cdot4!}\tilde{F}_{m}^{\;\;pqrs}\;
d^{I_{1}}\nabla_{[n}\epsilon_{pqrs]}^{\quad\;\;a}\;
\nabla_{a}Y^{I_{1}}
\quad =\quad 0 \;,\label{E3.1}
\\ \nonumber \\ 
&&\Bigg\{
\frac{1}{2}h_{\mu}^{I_{1}\;\;\mu}
-\frac{8}{15}\;\pi^{I_{1}}
\Bigg\}\;\nabla_{(m}\nabla_{n)}Y^{I_{1}}
\quad = \quad 0\;, 
\label{E3.2}
\\ \nonumber \\ 
&&\Bigg\{
\nabla^{\rho}h_{\rho}^{I_{5}}\;\Bigg\}
\nabla_{(m}Y_{n)}^{I_{5}}
\quad =\quad 0
\label{E3.3} \;, 
\end{eqnarray}

\begin{eqnarray}
&\Bigg\{&
\frac{1}{2}(\nabla^{2}_{x}+\nabla_{y}^2-5e^2)\;h^{I_{14}}
-\left(\frac{1}{4}\nabla_{x}^2\Phi
-\frac{1}{2}(\nabla\Phi)^2
+\frac{1}{48}(H^{\mu\nu\rho}H_{\mu\nu\rho}
+e^{2\Phi}F^{\mu\nu\rho}F_{\mu\nu\rho}
\right)\;h^{I_{14}}
\;\Bigg\}\;Y_{(mn)}^{I_{14}}
\nonumber \\
&&-\frac{1}{4!}\;
e^{2\Phi}\tilde{F}_{mpqr}^{\qquad a}\tilde{F}_{n}^{\;pqr\,b}\;h^{I_{14}}
\;Y_{(ab)}^{I_{14}}
\quad =\quad 0\;. 
\label{E3.4}
\end{eqnarray}\\

\begin{itemize}
\item Harmonic expansions of linearized 2-form equations 
(\ref{NS2.1}), (\ref{NS2.2}), (\ref{NS2.3}), 
(\ref{R2.1}), (\ref{R2.2}), and (\ref{R2.3}) :
\end{itemize}
\vspace*{-10mm}
\begin{eqnarray}
&\Bigg\{&
(Max+\nabla_{y}^2)\;b^{I_{1}}_{\mu\nu}
-2\nabla^{\rho}\Phi{\cal H}^{I_{1}}_{\rho\mu\nu}
+H_{\mu\nu\rho}\nabla^{\rho}c^{I_{1}}
+\nabla^{\rho}H_{\rho\mu\nu}\;c^{I_{1}}
\nonumber \\
&&
-\frac{9}{4}\nabla^{\rho}\phi^{I_{1}}H_{\rho\mu\nu}
-\left(\frac{5}{2}\nabla^{\rho}H_{\rho\mu\nu}
-5\nabla^{\rho}\Phi H_{\rho\mu\nu}
+\frac{1}{4}\tilde{F}_{\mu\nu\rho\sigma\tau}
F^{\rho\sigma\tau}
\right)\phi^{I_{1}}
\nonumber \\
&&
+\frac{2}{3}H_{\mu\nu\rho}\nabla^{\rho}\pi^{I_{1}}
+\left(\frac{1}{3}\nabla^{\rho}H_{\rho\mu\nu}
-\frac{2}{3}\nabla^{\rho}\Phi H_{\rho\mu\nu}
+\frac{1}{6}e^{2\Phi}\tilde{F}_{\mu\nu\rho\sigma\tau}
F^{\rho\sigma\tau}\right)\pi^{I_{1}}
\nonumber \\ 
&&
-H_{\sigma\mu\nu}(\nabla^{\rho}h^{I_{1}\;\sigma}_{\;\rho}
-\frac{1}{2}\nabla^{\sigma}h^{I_{1}\;\rho}_{\;\rho})
-H_{\rho\sigma\nu}\nabla^{\rho}h^{I_{1}\;\sigma}_{\;\mu}
-H_{\rho\mu\sigma}\nabla^{\rho}h^{I_{1}\;\sigma}_{\;\nu}
\nonumber \\
&& \hspace{-30mm} 
+\left(-\nabla^{\rho}H^{\sigma}_{\;\mu\nu}
+2\nabla^{\rho}\Phi H^{\sigma}_{\;\mu\nu}
-\frac{1}{2}e^{2\Phi}
\tilde{F}_{\mu\nu\lambda\tau}^{\quad\;\;\rho}
F^{\lambda\tau\sigma}
\right)h^{I_{1}}_{\rho\sigma}
+\frac{1}{3!}e^{2\Phi}\tilde{F}_{\mu\nu}^{\quad\rho\sigma\tau}
{\cal F}^{I_{1}}_{\rho\sigma\tau}
+\frac{1}{3!}e^{2\Phi}F^{\rho\sigma\tau}
\tilde{{\cal F}}^{I_{1}}_{\mu\nu\rho\sigma\tau}
\Bigg\}\;Y^{I_{1}} = 0  \nonumber \\ 
\label{(NSNS2.1)}
\end{eqnarray}

\begin{eqnarray}
&\Bigg\{&
(Max+\nabla_{y}^2)\;a^{I_{1}}_{\mu\nu}
-\frac{1}{4}F_{\rho\mu\nu}\nabla^{\rho}\phi^{I_{1}}
-\left(\frac{1}{2}\nabla^{\rho}F_{\rho\mu\nu}
+\frac{1}{4}\tilde{F}_{\mu\nu\rho\sigma\tau}
H^{\rho\sigma\tau}
\right)\phi^{I_{1}}
\nonumber \\
&&+H_{\mu\nu\rho}\nabla^{\rho}c^{I_{1}}
+\nabla^{\rho}H_{\rho\mu\nu}\;c^{I_{1}}
+\frac{2}{3}H_{\mu\nu\rho}\nabla^{\rho}\pi^{I_{1}}
+\left(\frac{1}{3}\nabla^{\rho}F_{\rho\mu\nu}
-\frac{1}{6}\tilde{F}_{\mu\nu\rho\sigma\tau}
H^{\rho\sigma\tau}\right)\pi^{I_{1}}
\nonumber \\
&&-F_{\sigma\mu\nu}(\nabla^{\rho}h^{I_{1}\;\sigma}_{\;\rho}
-\frac{1}{2}\nabla^{\sigma}h^{I_{1}\;\rho}_{\;\rho})
-F_{\rho\sigma\nu}\nabla^{\rho}h^{I_{1}\;\sigma}_{\;\mu}
-F_{\rho\mu\sigma}\nabla^{\rho}h^{I_{1}\;\sigma}_{\;\nu}
-\left(\nabla^{\rho}F^{\sigma}_{\;\mu\nu}
+\frac{1}{2}\tilde{F}_{\mu\nu\lambda\tau}^{\quad\;\;\rho}
H^{\lambda\tau\sigma}
\right)h^{I_{1}}_{\rho\sigma}
\nonumber \\
&&-\frac{1}{3!}\tilde{F}_{\mu\nu}^{\quad\rho\sigma\tau}
{\cal H}^{I_{1}}_{\rho\sigma\tau}
-\frac{1}{3!}H^{\rho\sigma\tau}
\tilde{{\cal F}}^{I_{1}}_{\mu\nu\rho\sigma\tau}
\Bigg\}\;Y^{I_{1}}
\quad =\quad 0\;,
\label{(RR2.1)}
\end{eqnarray}

\begin{eqnarray}
&&\Bigg\{
\nabla^{\rho}b_{\rho\mu}^{I_{1}}
-2(\nabla^{\rho}\Phi)\;b_{\rho\mu}^{I_{1}}
+\frac{1}{3!}e^{2\Phi}F^{\rho\sigma\lambda}\;
\tilde{D}_{\rho\sigma\lambda\mu}^{I_{1}}
\Bigg\}\;\nabla_{m}Y^{I_{1}}
\quad =\quad 0 \;,
\label{(NSNS2.2a)}
\\ \nonumber \\
&&\Bigg\{
\nabla^{\rho}a_{\rho\mu}^{I_{1}}
-\frac{1}{3!}H^{\rho\sigma\lambda}\;
\tilde{D}_{\rho\sigma\lambda\mu}^{I_{1}}
\Bigg\}\;\nabla_{m}Y^{I_{1}}
\quad =\quad 0 \;,
\label{(RR2.2a)}
\end{eqnarray}

\begin{eqnarray}
&&\Bigg\{
(Max+\nabla_{y}^2-4e^2)\;b_{\mu}^{I_{5}}
-2\nabla^{\rho}\Phi(2\partial_{[\rho}b^{I_{5}}_{\mu]})
+H_{\mu}^{\;\;\rho\sigma}\nabla_{\rho}h^{I_{5}}_{\;\sigma}
+\frac{1}{3!}e^{2\Phi}F^{\rho\sigma\tau}
\tilde{{\cal F}}^{I_{5}}_{\rho\sigma\tau\mu}
\Bigg\}\;Y_{m}^{I_{5}}
 = 0\;,
\label{NSNS2.2b} 
\nonumber \\ \\
&&\Bigg\{
(Max+\nabla_{y}^2-4e^2)\;a_{\mu}^{I_{5}}
+F_{\mu}^{\;\;\rho\sigma}\nabla_{\rho}h_{I_{5}\;\sigma}
-\frac{1}{3!}H^{\rho\sigma\tau}
\tilde{{\cal F}}^{I_{5}}_{\rho\sigma\tau\mu}
\Bigg\}\;Y_{m}^{I_{5}}
 =0\;,
\label{RR2.2b} 
\end{eqnarray}

\begin{eqnarray}
&\Bigg\{&
\nabla^{\rho}b^{I_{5}}_{\rho}
+2\nabla^{\rho}\Phi \;b^{I_{5}}_{\rho}
-\frac{1}{3!}e^{2\Phi}F^{\rho\sigma\tau}
\tilde{D}^{I_{5}}_{\rho\sigma\tau}
\Bigg\}\;2\;\nabla_{[m}Y^{I_{5}}_{n]}
\quad =\quad 0\;,
\label{NSNS2.3a} \\
&\Bigg\{&
\nabla^{\rho}a^{I_{5}}_{\rho}
+\frac{1}{3!}H^{\rho\sigma\tau}
\tilde{D}^{I_{5}}_{\rho\sigma\tau}
\Bigg\}\;2\;\nabla_{[m}Y^{I_{5}}_{n]}
\quad =\quad 0\;,
\label{RR2.3a} 
\end{eqnarray}

\begin{eqnarray}
&\Bigg\{&
(\nabla_{x}^2+\nabla_{y}^2-6e^2)\;b^{I_{10}}
-2\nabla^{\mu}\Phi\nabla_{\mu}b^{I_{10}}
+\frac{1}{3!}e^{2\Phi}F^{\mu\nu\rho}
\tilde{{\cal F}}^{I_{10}}_{\mu\nu\rho}
\Bigg\}\;Y^{I_{10}}_{[mn]}
\nonumber \\
&&+\frac{1}{3!}e^{2\Phi}\tilde{F}_{mn}^{\quad pqr}
b^{I_{10}}\;\nabla_{p}Y^{I_{10}}_{[qr]} \quad=\quad 0 \;,
\label{NSNS2.3b} 
\\  \nonumber \\
&\Bigg\{&
(\nabla_{x}^2+\nabla_{y}^2-6e^2)\;a^{I_{10}}
-\frac{1}{3!}H^{\mu\nu\rho}
\tilde{{\cal F}}^{I_{10}}_{\mu\nu\rho}
\Bigg\}\;Y^{I_{10}}_{[mn]}
-\frac{1}{3!}e^{2\Phi}\tilde{F}_{mn}^{\quad pqr}
a^{I_{10}}\;\nabla_{p}Y^{I_{10}}_{[qr]} \;=\;0 .
\label{RR2.3b} 
\end{eqnarray}
\\

\begin{itemize}
\item Harmonic expansions of 
linearized self-dual equations 
(\ref{SD1}), (\ref{SD2}) and (\ref{SD3}) :
\end{itemize}
\begin{eqnarray}
&&\hspace{-10mm}
\Bigg\{
\tilde{{\cal F}}_{\mu\nu\rho\sigma\lambda}^{I_{1}}
-\epsilon_{\mu\nu\rho\sigma\lambda}
\;\nabla_{y}^2\;d^{I_{1}}
-\frac{1}{2\cdot5!}\;\epsilon_{\mu\nu\rho\sigma\lambda}
^{\qquad mnlkj}
\tilde{F}_{mnlkj}\left(
h_{\rho}^{I_{1}\;\;\rho}
-\frac{8}{3}\;\pi^{I_{1}}\right)\;
\Bigg \}\;Y^{I_{1}}
\quad =\quad 0 \;,
\label{SD1.1} 
\\ \nonumber \\
&&\hspace{-10mm}
\Bigg\{
\tilde{D}_{\mu\nu\rho\sigma}^{I_{1}}
+\epsilon_{\mu\nu\rho\sigma\lambda}\;
\nabla^{\lambda}\;d^{I_{1}}
\Bigg\}\;\nabla_{m}Y^{I}
\quad =\quad 0\;,
\label{SD2.1}
\\ \nonumber \\
&&\hspace{-10mm}
\Bigg\{
\tilde{{\cal F}}_{\mu\nu\rho\sigma}^{I_{5}}
+\;\epsilon_{\mu\nu\rho\sigma}^{\quad \lambda}
\;d_{\lambda}^{I_{5}}\;\nabla_{y}^2
\Bigg\}\;Y_{m}^{I_{5}}(y)
+\!\!\Bigg\{
\epsilon_{\mu\nu\rho\sigma}^{\qquad \lambda}
\;d^{I_{5}}_{\lambda}\;R_{mn}^{\quad ni}
+\frac{1}{4!}\;
\epsilon_{\mu\nu\rho\sigma m}^{\qquad \;\;nlkj\lambda}
\tilde{F}_{nlkj}^{\quad\;\; i}\;h_{\lambda}^{I_{5}}
\Bigg\}\;Y_{i}^{I_{5}}
= 0 ,
\label{SD2.2}
\nonumber \\  \\
&&\hspace{-10mm}
\Bigg\{
\tilde{D}_{\mu\nu\rho}^{I_{5}}
+\;(\nabla_{\sigma}d_{\lambda}^{I_{5}}-\nabla_{\lambda}d_{\sigma}^{I_{5}})
\;\epsilon_{\mu\nu\rho}^{\quad \sigma\lambda}
\Bigg\}\;\nabla_{[m}Y_{n]}^{I_{5}}
\quad = \quad 0\;,
\label{SD3.1}
\\ \nonumber \\
&&\hspace{-10mm}
\Bigg\{
\tilde{{\cal F}}^{I_{10}}_{\mu\nu\rho}
-\frac{10}{5!}\;
\epsilon_{\mu\nu\rho mn}^{\qquad \;\;lkj\sigma\lambda}\;
\tilde{D}_{\sigma\lambda}^{I_{10}}
\;\Bigg\}\;\nabla_{[l}Y^{I_{10}}_{kj]}
\quad =\quad 0\;.
\label{SD3.2}
\end{eqnarray}

\newpage

\newpage
\def\refname{References}

\end{document}